\documentclass[apj,tighten]{emulateapj2} % Fix for color tables
\usepackage{natbib}
\usepackage[usenames]{color}
\definecolor{red}{rgb}{0.75,0,0}
\definecolor{green}{rgb}{0,0.5,0}
\definecolor{blue}{rgb}{0,0,0.75}
\definecolor{Yellow}{rgb}{1, 0.75, 0}
\usepackage[ colorlinks=true, % false for boxed links
             urlcolor=blue,
             filecolor=blue,
             linkcolor=red,
             citecolor=blue,
             pdftitle={The Morphologies of Green Galaxies},
             pdfauthor={Alexander Mendez},
             bookmarksopen=true]{hyperref}
\usepackage{colortbl}
\usepackage{array}
\usepackage{amssymb,amsmath} 
\usepackage[figuresleft]{rotating}

\begin{document}
\newcommand{\halfwidth}{9cm}
\newcommand{\fullwidth}{15cm}
\newcommand{\mtwe}{\ensuremath{M_{20}}}
\newcommand{\mtheta}{\ensuremath{M\arcmin_{20}}}
\newcommand{\gtheta}{\ensuremath{G\arcmin}}
\newcommand{\ugriz}{\ensuremath{ugriz } }
\newcommand{\galex}{\emph{GALEX} }
\newcommand{\hst}{\emph{HST}}
\newcommand{\bt}{\ensuremath{B/T}}
\newcommand{\rp}{\ensuremath{r_p}}
\newcommand{\nuv}{\ensuremath{_{ \tt NUV}} }
\newcommand{\cas}{\ensuremath{{CAS}}}
\newcommand{\full}{{\tt Full} }
\newcommand{\limited}{{\tt Limited} }
\newcommand{\stellarmass}{{\tt Stellar~Mass} }
\newcommand{\rgb}{{\tt RGB} }

\title{AEGIS: The Morphologies of Green Galaxies at $0.4<z<1.2$}
\author{Alexander J. Mendez \altaffilmark{1},
        Alison L. Coil\altaffilmark{1,2},
        Jennifer Lotz\altaffilmark{3},
        Samir Salim\altaffilmark{3},
        John Moustakas\altaffilmark{1},
        Luc Simard\altaffilmark{4}}
\altaffiltext{1}{Department of Physics, University of California San Diego, 
                 9500 Gilman Dr., La Jolla, CA 92093}
\altaffiltext{2}{Alfred P. Sloan Foundation Fellow}
\altaffiltext{3}{National Optical Astronomical Observatory, 
                 950 North Cherry Ave., Tucson, AZ 85719}
\altaffiltext{4}{National Research Council of Canada, 
                 Herzberg Institute of Astrophysics, 
                 5071 West Saanich Road, Victoria, British Columbia, V9E 2E7, Canada}

\begin{abstract} 
We present quantitative morphologies of $\sim$300 galaxies in the
optically-defined green valley at $0.4<z<1.2$, in order to constrain the
mechanism(s) responsible for quenching star formation in the bulk of this
population. The sample is selected from galaxies in the All-Wavelength Extended
Groth Strip International Survey (AEGIS). While the green valley is defined
using optical U-B colors, we find that using a green valley sample defined using
NUV-R colors does not change the results. Using \hst/ACS imaging, we study
several quantitative morphological parameters including CAS, $\bt$ from GIM2D,
and Gini/$\mtwe$. We find that the green galaxy population is intermediate
between the red and blue galaxy populations in terms of concentration,
asymmetry, and morphological type and merger fraction estimated using
Gini/$\mtwe$. We find that most green galaxies are {\it not} classified as
mergers; in fact, the merger fraction in the green valley is {\it lower} than in
the blue cloud. We show that at a given stellar mass, green galaxies have higher
concentration values than blue galaxies and lower concentration values than red
galaxies. Additionally, we find that 12\% of green galaxies have $\bt=0$ and 21\% with
$\bt\le0.05$. Our results show that green galaxies are generally massive ($M_{*}\sim
10^{10.5}$ $M_{\sun}$) disk galaxies with high concentrations. We conclude that
major mergers are likely not the sole mechanism responsible for quenching star
formation in this population and that either other external processes or
internal secular processes play an important role both in driving gas towards
the center of these galaxies and in quenching star formation.
\end{abstract}

%%%%%%%%%%%%%%%%%%%%%%%%%%%%%%%%%%%%%%%%%%%%%%%%%%%%%%%%%%%%%%%%%%%%%%%%%%%%%%%%
\section{Introduction} \label{sec:introduction}
%%%%%%%%%%%%%%%%%%%%%%%%%%%%%%%%%%%%%%%%%%%%%%%%%%%%%%%%%%%%%%%%%%%%%%%%%%%%%%%%
The wealth of data generated by local large redshift surveys such as the
\emph{Sloan Digital Sky Survey} \citep[SDSS,][]{York00} and the \emph{Two-Degree
Field Galaxy Redshift Survey} \citep[2dFGRS,][]{Colless01} have greatly advanced
our understanding of galaxy properties at $z\sim0.1$. These surveys have clearly
established that the local galaxy population exhibits a bimodal distribution in
terms of optical color \citep[e.g.,][]{Strateva01,Blanton03}, UV-optical color
\citep{Salim07}, the 4000\AA~ break $D_n-4000$ \citep{Kauffmann03}, and spectral
type \citep{Madgwick02}. In optical color-magnitude diagrams (CMDs) galaxies
predominantly lie along either the ``red sequence'', which is dominated by
quiescent, non-star-forming, early-type, bulge-dominated galaxies
\citep[e.g.,][]{Zhu10, Blanton09}, or in the ``blue cloud,'' characterized by
star-forming, late-type, disk-dominated galaxies.

Deeper redshift surveys that probe galaxies at an earlier stage of evolution
have shown that this bimodality exists at least to $z\sim2$
\citep[e.g.,][]{Willmer06,Faber07,Kriek08,Williams09}. The location of the
bimodality minimum is bluer at $z\sim1$ by approximately 0.1 mag, as both blue
and red galaxies were bluer in the past \citep{Blanton06}. Results from the
COMBO-17 \citep{Bell04} and NOAO Deep Wide-Field Survey \citep{Brown07}
photometric redshift surveys and the DEEP2 spectroscopic redshift survey
\citep{Faber07} show that the red sequence has grown in mass by a factor of 2-4
since $z\sim1$, while the number density of galaxies in the blue cloud has
remained roughly constant \citep{Bell04,Brown07,Faber07}. The large influx of
red sequence galaxies brighter than $L_*$ at $z \leq 0.7$ is dominated by
spheroidal systems \citep{Blanton03,Bell06,Weiner05,Scarlata07}, but red
disk-dominated galaxies are more common at fainter magnitudes \citep{Brown07}.

Recent observational studies investigating galaxies in the minimum of the
optical color bimodality, so-called ``green valley'' galaxies, have begun to
probe the nature of this population. \citet{Baldry04} model the u-r color
distribution of SDSS galaxies and find that it can be fit as the sum of separate
Gaussian distributions for the red and blue populations, implying that green
valley galaxies are not necessarily a distinct population. However,
\citet{Wyder07} use UV-optical colors to more clearly separate the star-forming
and quiescent populations and show that there is an excess of galaxies in the
green valley galaxy population at $z\sim0.1$; thus green galaxies may not be a
simple mix of blue and red galaxies. Additionally, \citet{Salim09} show in a
comparison of the specific star formation rate (SSFR) versus rest-frame
UV-optical color for low redshift galaxies that there is a smooth transition
from high to low SSFR as color increases; galaxies at intermediate colors do not
have a SSFR distribution that encompasses the values seen for both red and blue
galaxies and thus appear to be a transition population.

The observed bimodality of galaxy colors reflects the fact that galaxies are
either actively forming stars and are optically blue, are not forming stars and
are optically red, or are dusty. After a galaxy stops forming stars, it should
move from the blue cloud to the red sequence; star formation quenching could
therefore explain the observed evolutionary trends. From the observed number
densities and clustering properties of galaxies as a function of color, the
implied timescale for movement to the red sequence after the quenching of star
formation must be short, on the order of $\sim1$ Gyr, otherwise there would be
no observed bimodality \citep{Faber07,Martin07,Tinker10}. The short time scale
is also a natural consequence of stellar population models; without star
formation blue galaxies turn red in $\sim1$ Gyr \citep{Bruzual03}.

It is not yet known what physical mechanism or mechanisms cause these galaxies
to stop forming stars. A variety of star formation quenching mechanisms have
been proposed. It has long been suggested that major mergers are the dominant
mechanism responsible for converting blue, star-forming, spiral galaxies into
red, quiescent, elliptical galaxies. Simulations show that major mergers
randomize the orbits of stars within a galaxy and can change the overall galaxy
morphology from a disk into a bulge \citep[e.g.][]{Toomre72,Toomre77,Barnes92,
Hernquist92,Hernquist93,Naab03,Cox06a}. During the merger, gas is funneled to
the center of the remnant, resulting in a burst of star formation that consumes,
expels and/or heats some fraction of the available gas through shocks or
feedback from supernovae \citep[e.g.][]{Barnes96,Springel05b,Robertson06,Cox06b,
Steinmetz02}.

Merger-induced starbursts alone are likely insufficient to fully quench star
formation or to consume \emph{all} of the available gas, particularly for gas
rich mergers at high redshift; additional quenching or gas removal is needed
\citep{Dekel06,Birnboim07}. \citet{Cox08} show that in smoothed particle
hydrodynamic simulations the starburst efficiency in merger-induced starbursts
in recent simulations is lower than what was found previously. Their results
suggest that these starbursts will not fully consume or eject all of the gas in
the system. Additionally, \citet{Lotz08b} use similar simulations to show that
major merger remnants have enhanced SFR for $\sim$1 Gyr \emph{after} the
coalescence of the nuclei, indicating that additional gas removal mechanisms are
required. Residual star formation in the remnant can prevent the galaxy from
having red colors characteristic of elliptical galaxies \citep{Springel05a}. In
order for the green valley to exist, star formation must be quenched on
relatively short time scales, of order $\sim$1 Gyr. If instead the residual star
formation declines slowly over a Hubble time \citep{Mihos94}, the remnant would
gradually transition to the red sequence, removing the green valley distinction
between the red and blue galaxy populations.

Further, simulations show that cold gas accretion from the intergalactic medium
can also feed star formation in galaxies, particularly at high redshift
\citep{Dekel06,Birnboim07,Keres05,Keres09a,Brooks09}. In these simulations,
dense filamentary gas collapses to forms cold clouds which are stable against
shocks and can penetrate to the centers of dark matter halos and accrete onto
galaxies. This cold gas inflow further necessitates additional quenching
mechanisms, as red and dead galaxies must remain quenched for the majority of
cosmic time. Shock heating of the gas as it falls into a more massive dark
matter halo limits the cold gas accretion from the intergalactic medium. For
halos with masses above $\sim10^{12}$ $M_{\sun}$, the infalling gas is heated to
such a temperature that the cooling time is longer than the Hubble time, so that
it cannot radiatively cool, thereby forming a halo of hot gas \citep{Birnboim03,
Keres05,Dekel06}. \citet{Johansson09} show that gravitational heating may also
be an important mechanism in massive halos, through the release of potential
energy from infalling stellar clumps.

In addition to shock heating of infalling gas, \citet{Croton06} has incorporated
``radio-mode'' Active galactic nucleus(AGN) feedback into simulations, in which
an AGN heats gas in massive structures, such as galaxy groups and clusters to
limit intergalactic gas from being able fall into the galaxy. This form of AGN
feedback is observationally supported by X-ray imaging of evacuated cavities
around massive galaxies in the centers of clusters \citep{McNamara00,McNamara01,
McNamara07}. Quasar mode AGN feedback, invoked in many current galaxy evolution
models \citep{Hopkins06,Dave01}, could potentially limit the amount of cold
gas available for star formation, but there is little direct observational
evidence for this picture.

Interestingly, the rate of AGN detection is high in green valley galaxies,
whether AGN are selected by deep X-ray surveys \citep{Nandra07,Coil09,Hickox08,
Schawinski09,Cardamone10} or by optical line-ratio diagnostics \citep{Salim07}.
\citet{Nandra07} find that many X-ray AGN host galaxies are green defined using
(U-B) colors, while \citet{Pierce10} find the same trend using (NUV-R) colors.
\citet{Coil08} show that in coadded spectra of DEEP2 galaxies, the average
spectrum of green galaxies at $z\sim1$ is not a simple mix of the average
spectra of blue and red galaxies but instead shows line ratios indicative of
enhanced AGN activity. Additionally, \citet{Bundy08} find that the star
formation quenching rate from $z\sim1$ to today is consistent with the AGN
``trigger'' rate. It remains unclear, however, whether the presence of an AGN in
a green valley galaxy is directly related to the star formation quenching
process.

On the observational side, much recent work shows that secular effects may also
play a crucial role in establishing the color bimodality. The majority of nearby
disk galaxies ($\sim75\%$) are found to have a stellar bar \citep{Eskridge00,
Menendez07}, and the bar fraction remains high to $z\sim1$ \citep{Sheth03,
Elmegreen04,Jogee04}. The large non-axially-symmetric potential of bar galaxies
can induce a large-scale inflow of stars and gas \citep{Sellwood93, Sheth05},
and studies of gas kinematics in the bar indicate that molecular gas flows
inward along the bar dust lanes \citep{Downes96, Regan99,Sheth00, Sheth02}.
Additionally, \citet{Lopez10b} find that minor mergers can not fully account for
the mass growth of galaxies, suggesting that secular processes are needed to
generate a bulge-dominated population. These bulge-dominated galaxies may be
populating early-type spirals such as those observed by \citet{Bundy10}.
Interestingly, \citet{Masters10b} and \citet{Cameron10} find that early-type
spirals have higher bar fractions than late-types spirals. At $z\sim1$
\citet{Oesch10} and \citet{Lopez10a} find that major mergers are not common
enough to explain the late- to early-type transition and suggest that either
minor mergers or secular processes are needed.

Other potential quenching mechanisms need to have enough energy to halt star
formation and continue to keep star formation quenched over a Hubble time. Many
of the proposed mechanisms either heat the gas such that it cannot collapse to
form stars \citep[e.g. AGN feedback,][]{Bower06,Croton06,Hopkins06,Kang06,
Springel05a}, remove cold gas from the galaxy as it falls into a halo
\citep[e.g. ram-pressure stripping][]{Gunn72,Kimm09,Quilis00,Hester06,
Farouki80,Moore96,Abadi99}, or remove hot, diffuse, gas from a satellite galaxy
\citep[e.g. ``strangulation''][]{Tinsley80,Balogh00}. Tidal stripping of gas
along the orbit of a satellite galaxy both removes gas and causes the galaxy to
become more concentrated \citep{Bosch08}, unlike ram-pressure stripping and
strangulation, which mainly influence the gas in the galaxy as opposed to the
stars. Additionally, gravitational interactions can greatly alter the morphology
and gas content of a galaxy from the cumulative effect of many high-speed
impulsive encounters, known as ``harassment'' \citep{Farouki81,Moore96}.

The variety of proposed star formation quenching mechanisms mentioned above
should have different morphological consequences. Therefore, in principle, one
should be able to study the morphologies of green valley galaxies to constrain
the dominant quenching mechanism. For example, parametric measures such as
galaxy asymmetry can identify disturbed morphologies caused by major mergers
\citep{Conselice00}. In addition, the Gini coefficient ($G$) and second-order
moment of light ($\mtwe$) parameters can be used to identify both major and
minor mergers \citep{Lotz10}. Gas stripping should lead to galaxies with
truncated disks \citep{Moore96}, while enhanced star formation due to harassment
or other environmental interactions should lead to localized luminous areas,
increasing the clumpiness (and decreasing the smoothness) of galaxies
\citep{Moore98}.

Previous studies of morphology and color have found that while most red galaxies
have early-type, bulge-dominated morphologies, morphologically intermediate-type
galaxies (type Sa-Sbc) are scattered throughout the CMD, including the red
sequence \citep[e.g.,][]{Ball06,Driver06,Bell03,Blanton09,Pannella09}. It is not
yet clear if the majority of galaxies on the red sequence were already
spheroidal when they first joined the red sequence or if they were still
disk-dominated and later became spheroidal through mergers with other red
galaxies, or if a combination of both processes occurs \citep{Faber07}. Galaxies
could potentially also move from the red sequence to the blue cloud, for example
as the result of a merger between a blue and red galaxy. However, the bulk of
the movement must be from the blue cloud to the red sequence to explain the
observed mass and number density evolution on the red sequence.

In this paper we investigate the morphologies of green valley galaxies at
$z\sim1$ to a) compare their morphological distributions to those of blue and
red galaxies at the same redshift, to constrain the dominant star formation
quenching mechanism(s) at work, and b) test whether the morphological
distribution of green galaxies is consistent with being a simple mix of red and
blue galaxies, or whether green galaxies have a distinct morphological makeup.
We use a sample of galaxies at $0.4<z<1.2$ from the All-Wavelength Extended
Groth Strip International Survey (AEGIS) \citep{Davis07}, combining \hst/ACS
imaging with DEEP2 spectroscopic and CFHTLS photometric redshifts to measure
various quantitative morphological parameters.

The outline of the paper is as follows: In \S\ref{sec:data} we present the AEGIS
datasets used here. In \S\ref{sec:sample} we define the red, green and blue
samples used here. In \S\ref{sec:parameters} we discuss the different
morphological parameters measured. In \S\ref{sec:results} we present results on
the morphologies of green galaxies compared to the red and blue galaxies at the
same redshift. In \S\ref{sec:purpleresults} we perform statistical tests with
control samples of blue and red galaxies to determine whether the green galaxy
population has a distinct morphological distribution. Finally, we discuss and
summarize the results in \S\ref{sec:discussion}. Absolute magnitudes given in
this paper are in the AB system and are $M_B-5$ ${\rm log}(h)$ with $h=0.7$,
which for the remainder of the paper we denote as $M_B$. We assume the standard
flat $\Lambda$CDM model with $\Omega_m=0.3$ and $\Omega_\Lambda=0.7$.

%%%%%%%%%%%%%%%%%%%%%%%%%%%%%%%%%%%%%%%%%%%%%%%%%%%%%%%%%%%%%%%%%%%%%%%%%%%%%%%%
\section{Data} \label{sec:data}
%%%%%%%%%%%%%%%%%%%%%%%%%%%%%%%%%%%%%%%%%%%%%%%%%%%%%%%%%%%%%%%%%%%%%%%%%%%%%%%%
To quantify the morphological distribution of green valley galaxies, we require
a large, complete parent sample of galaxies with accurate redshifts, rest-frame
colors and magnitudes, and high-resolution imaging. As galaxies in the green
valley have a lower space density than either blue or red galaxies, we require a
large parent sample to measure the bivariate distribution of morphological
parameters with sufficient objects per bin. We use data from the AEGIS survey
\citep{Davis07}, which covers the Extended Groth Strip (EGS) and contains DEEP2
spectroscopic and CFHTLS photometric redshifts. We use \hst/ACS imaging and
CFHTLS photometry to define a complete galaxy sample large enough to allow us to
quantify the joint morphological properties of green galaxies. We further use
\galex photometry in the field to test the effects of excluding green galaxies
that are likely dusty, star-forming blue galaxies, as opposed to true
transitionary galaxies. Table~\ref{table:data} contains a summary of the
datasets and sample sizes.

  % Data Table: ----------------------------------------------------------------
  \begin{deluxetable*}{llccl}
      \tablecolumns{5}
      \tablewidth{0pt}
      \tablecaption{Summary of Data Sets \label{table:data}}
      %snip%
		  \tablehead{Survey             & \colhead{Wavelengths/Bands} & \colhead{Sample Size \tablenotemark{a}} &    \colhead{Survey Limit}  }
		  \startdata
		               \hst/ACS Imaging &                         $V$ &                                    2437 &           28.75($V_{AB}$) \\
		                     \textrm{"} &                         $I$ &                                    2437 &           28.10($I_{AB}$) \\
		  DEEP2 Spectroscopic Redshifts &                6400-9100\AA &                                    1220 &           24.1 ($R_{AB}$) \\
		   CFHTLS Photometric Redshifts &                      \ugriz &                                    2324 &             25.($i_{AB}$) \\
		    \galex Deep Imaging Survey  &                       $NUV$ &                                    1021 &                  26.5(AB) \\
		                     \textrm{"} &                       $FUV$ &                                     345 &                    25(AB) \\
		  CFHT Legacy Survey Photometry &                      \ugriz &                                    2437 &              $\sim$27(AB) \\
      %snip%
      \tablenotetext{a}{Number of galaxies with redshift with $0.4<z<1.2$,
      average signal-to-noise per pixel greater than 4, Petrosian radius
      $\rp>0.3\arcsec$, and $M_B>-18.$}
  \end{deluxetable*}

  % Areal Map of the Samples: --------------------------------------------------
  \begin{figure}
    \epsscale{1.2}
    \plotone{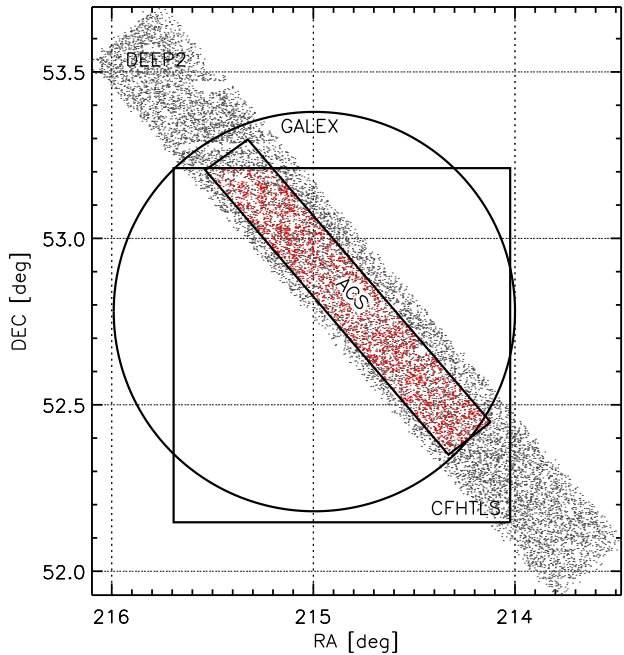}
    \caption{Areal map of the various relevant datasets in the EGS. Individual
    DEEP2 galaxies are shown in black points to a limit of to $R_{\rm AB}<24.1$
    and our sample is shown in red. Here we use galaxies that lie within the ACS
    region of the EGS and have secure DEEP2 spectroscopic redshifts (grey
    points) or good CFHTLS photometric redshifts (square region), with
    additional cuts (see Table~\ref{table:data} for sample sizes and depths). A
    single $1\deg$ deep \galex pointing also overlaps a majority of the region.}
    \label{fig:skydata}
  \end{figure}

  \subsection{\hst/ACS Imaging} High-resolution images from the Hubble Space
  Telescope (\hst) Advanced Camera for Surveys (ACS) were obtained from the
  AEGIS survey as part of the GO Program, 10134 PI: M. Davis \citep{Davis07}.
  The EGS was imaged in both the $V$ (F606 W, 2260 sec) and $I$ (F814W, 2100
  sec) bands in a $\sim$10\farcm1 by 67\farcm~ strip along the field. For
  details about the ACS imaging and reduction see \citet{Davis07} and
  \citet{Lotz08a}. The 5$\sigma$ limiting magnitudes for a point source are
  $V_{\rm AB}=28.75$ and $I_{\rm AB}=28.10$.

  \subsection{CFHTLS \ugriz Photometry} We use CFHTLS T0004 release \ugriz
  photometry \citep{Ilbert06} to calculate rest-frame colors and magnitudes for
  each galaxy. The CFHTLS Deep Field 3 is a $1\arcdeg\times1\arcdeg$ field that
  covers the majority of the ACS region (see Figure~\ref{fig:skydata}). We use
  CFHTLS photometry, flux limited to $i_{\rm AB}<25$, for objects in the ACS
  region, to calculate rest frame magnitudes. There are 16,450 objects in the
  ACS footprint with CFHTLS \ugriz photometry.

  \subsection{$\galex$ NUV/FUV Photometry} We use near-UV (NUV) and far-UV (FUV)
  data obtained from the single 1\fdg2 diameter pointing of the central region
  of the EGS, taken with the \emph{Galaxy Evolution Explorer}
  \citep[$\galex$;][]{Martin05, Morrissey07}. The 237 ks of NUV imaging and 120
  ks of FUV imaging data are part of the third data release (GR3)
  \citep{Zamojski07}. We use deblended $\galex$ photometry from \citet{Salim09},
  which use an expectation maximization (EM) method to measure fluxes using a
  Bayesian deblending technique \citep{Guillaume06} and optical $u$-band priors
  \citep[see][ for details]{Salim09}. This method overcomes issues of blending
  and source confusion due to the 4$\arcsec$-5$\arcsec$ spatial resolution
  (FWHM) of $\galex$ \citep{Morrissey07}. NUV detections are measured to a flux
  limit of $NUV<26.5$, and a FUV flux limit of $FUV<25$.

  \subsection{Spectroscopic and Photometric Redshifts} The DEEP2 redshift survey
  provides spectroscopic redshifts to $R_{\rm AB}<24.1$ in AEGIS; see
  \citet{Davis07} for details. Here we use only redshifts between $0.4<z<1.2$
  with a confidence greater than 95\% $(z_{\rm Quality} \geq 3)$. With this
  quality cut, there are 2,885 galaxies with spectroscopic redshifts within the
  ACS region.

  We additionally use photometric redshifts to both increase our sample size and
  allow us to probe fainter flux limits. We use CFHTLS T0003 release photometric
  redshifts from \cite{Ilbert06}, derived from $\ugriz$ imaging covering the
  central $1\arcdeg\times1\arcdeg$ of the field. We remove from the sample all
  galaxies that are below our flux limit of $i_{AB}=25$, that have large
  photometric errors, or that have a 5\% or greater probability of being at a
  different redshift. We limit the sample to the redshift range $0.4<z<1.2$. At
  lower redshifts the volume probed is small, while the upper limit ensures that
  the ACS imaging samples rest-frame optical morphologies, thereby minimizing
  rest-frame wavelength-dependent morphology biases. Combined with DEEP2
  redshifts, this results in a sample of galaxies with redshifts in the range
  $0.4<z<1.2$ with a median redshift of $0.75$.

  Using galaxies that have both DEEP2 spectroscopic and CFHTLS photometric
  redshifts, we are able to test the redshift precision of the photometric
  redshifts. Within the redshift range $0.4<z_{spec}<1.2$, there are 3,306
  galaxies in both catalogs. Among these, 4.2\% are catastrophic outliers,
  defined as having $|{\Delta}z|/{(1+z_{spec})} > 0.15$. Excluding catastrophic
  errors, the photometric redshifts have an accuracy of
  $\sigma_{{\Delta}z/(1+z_{spec}}) = 0.038 $, with a normalized median absolute
  deviation: $\eta = 1.48$ and a median $[|{\Delta}z|/(1+z_{spec})] = 3.1\%$.
  See \citet{Ilbert06} for a full discussion of the photometric redshifts.

  Table~\ref{table:data} contains a summary of the datasets, sample sizes, and
  depths for the parent sample. The completeness of our parent sample depends on
  the detection limits of the ACS images and the completeness of the DEEP2 and
  CFHTLS redshift catalogs, which may depend on color and magnitude. The DEEP2
  spectroscopic targeting selection excludes objects with a surface brightness
  fainter than $\mu_R \sim 26.5$ \citep{Davis07}, whereas the CFHTLS photometric
  redshift catalog has no strong selection against low surface brightness
  objects as compared to the ACS detections \citep{Lotz08a}. Due to the
  detection in DEEP2 spectra of emission lines in blue star-forming galaxies and
  absorption features in older, red galaxies and CFHTLS measurements of spectral
  breaks in all galaxy populations, the redshift catalogs are not strongly
  biased against galaxies with either red or blue colors \citep{Lotz08a}. The
  dominate selection effect is due to the ACS detection limits of the morphology
  measurements. The ACS surface brightness detection limit of $\mu\sim24.7$
  results in the lowest surface brightness galaxies, which are likely to be
  blue, missing from the morphology catalog \citep{Lotz08a}. For our statistical
  tests, we ensure to always match the stellar mass distributions of the
  different comparison samples, which minimized the selection effect of missing
  the lowest surface-brightness blue galaxies.

  \subsection{K-Corrections and Stellar Masses} We use the CFHTLS \ugriz
  photometry and $\galex$ NUV/FUV magnitudes to calculate K-corrections
  \citep{Blanton07}. We convert the observed fluxes to rest-frame absolute
  $\galex$ NUV and $UBVR$ magnitudes in the Johnson-Morgan System.

  We derive stellar masses for galaxies in our sample following
  \citet{Weiner09}, who show that reasonably robust stellar masses at these
  redshifts can be inferred from the rest-frame $M_B$ and (U-B) colors,
  following color-M/L ratio relations \citep{Bell01}. Comparing these stellar
  masses to those calculated by \citet{Salim09}, who fit spectral energy
  distributions (SEDs) for AEGIS galaxies using up to eight bands of photometry
  from the UV to the near-Infrared, we find a good agreement, with an offset of
  0.13 dex and a scatter of 0.2 dex. The IMF used here and by \citet{Weiner09}
  is a ``Diet Salpeter'' IMF \citep{Bell03}.

%%%%%%%%%%%%%%%%%%%%%%%%%%%%%%%%%%%%%%%%%%%%%%%%%%%%%%%%%%%%%%%%%%%%%%%%%%%%%%%%
\section{Galaxy Sample Definitions} \label{sec:sample}
\begin{figure}
  \epsscale{1.2}
  \plotone{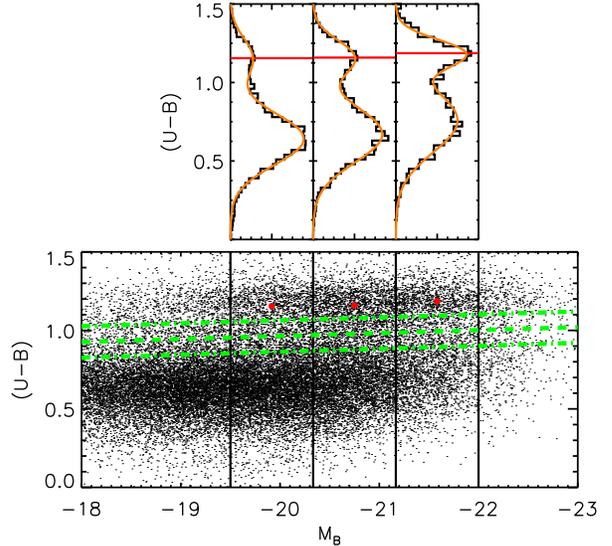}
  \caption{Color-magnitude diagram for AEGIS galaxies at $0.4<z<1.2$ and the
  definition of red, green, and blue galaxies. The upper panel shows double
  Gaussian fits (orange) to the (U-B) color distributions (black) for three
  magnitude bins between $M_B=$-19.5 and -23.0. We normalize the peak of the
  distribution in each magnitude bin to unity. We do not fit the
  magnitude-dependance of the red sequence below $M_B$=-19.5 or above $M_B$=-22
  due to the lower numbers of red galaxies. The red lines show the peaks of the
  fit, used to fit a linear magnitude-dependent slope to the red sequence. This
  same slope is used to define the center of the green valley. The lower panel
  shows the color-magnitude diagram for galaxies with $0.4<z<1.2$. Red points
  mark the locations of the peak of the red sequence for each magnitude bin, and
  the green dashed lines show the center and width of the green valley as
  defined by Equation~\ref{eqn:green}.}
  \label{fig:doublegauss}
\end{figure}

%%%%%%%%%%%%%%%%%%%%%%%%%%%%%%%%%%%%%%%%%%%%%%%%%%%%%%%%%%%%%%%%%%%%%%%%%%%%%%%%
We use rest-frame optical (U-B) colors to define our samples. As discussed
below, we have tested that our results do not change if we instead use (NUV-R)
colors, which more clearly separate star forming and quiescent galaxies. This
shows that our results are robust against contamination in the green valley by
dusty star forming galaxies. We define the rest-frame red, green, and blue
galaxy color samples by first locating the magnitude-dependent minimum of the
color bimodality. We use the well-defined magnitude-dependent slope derived
using the red sequence and then solve for a color offset to locate the minimum
of the green valley. To measure the slope of the red sequence, we fit a double
Gaussian to the observed (U-B) color distribution in three magnitude bins (shown
in the upper panel of Fig~\ref{fig:doublegauss}) and fit a linear
color-magnitude relation to the maximum of the Gaussian fit to the red galaxies
(shown as red dots in the lower panel of Figure~\ref{fig:doublegauss}). We then
fit for an offset of this line to match the minimum in the observed
color-magnitude diagram. The resulting definition of the center of the green
valley is

\begin{equation}\label{eqn:green}
		(U-B) = -0.0189(M_B + 19.79) + 0.96
\end{equation}

\noindent where the AB magnitude offset is given at the median of the parent
sample.

% Double Gaussian Fit: -------------------------------------------------------

% ColorCut Diagram: ----------------------------------------------------------
\begin{figure}
  \epsscale{1.2}
  \plotone{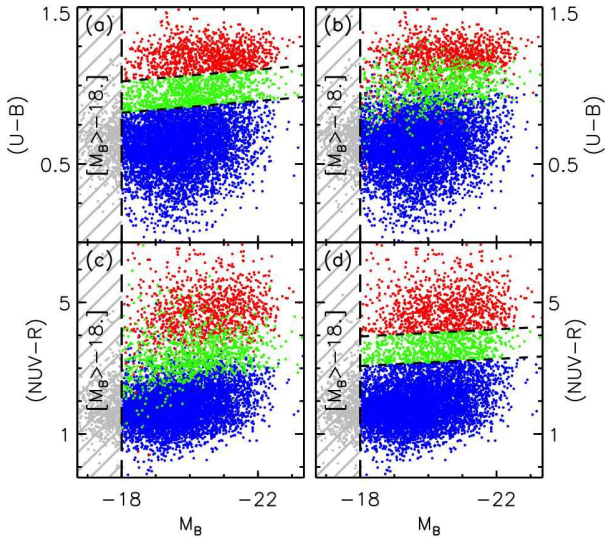}
  \caption{Comparison of green valley sample selection in different
  color-magnitude spaces. (a) Upper left panel: Rest-frame (U-B) color versus
  $M_B$ magnitude. Green valley galaxies are defined in this space using a
  tilted color cut (see Equation~\ref{eqn:green}). This is the sample
  definition used throughout the paper. We define the green valley to be within
  $\delta$(U-B)=0.1 of the minimum of the (U-B) color bimodality. Galaxies are
  color-coded here by whether they are defined to be red, green, or blue in this
  space. (c) Lower left panel: Rest-frame (NUV-R) versus $M_B$ for the same
  galaxies, color-coded by their definition in the panel above. (d) Lower right
  panel: We present an alternative definition of green valley galaxies using the
  observed bimodality in rest-frame (NUV-R) versus $M_B$ space. Galaxies here
  are color-coded by their definition in this space (see text for details). (b)
  Upper right panel: The same galaxies, color-coded by their definition in the
  panel below, are shown here in rest-frame (U-B) versus $M_B$ space. This
  figure shows that the selection of green valley galaxies at $0.4<z<1.2$ is
  very similar using either (U-B) or (NUV-R) colors.}
  \label{fig:colorcut}
\end{figure}

We define the green valley to have a width of $\delta (U-B)=0.1$ about the
minimum line, with red galaxies having redder colors and blue galaxies having
bluer colors. The width of the green valley is relatively arbitrary when using
optical colors; our choice follows \citet{Coil08}. See the dot-dashed lines in
Figure~\ref{fig:doublegauss} for the color sample definition. From our parent
sample of 2,437 galaxies, with this color-magnitude cut there is a total of 571
red, 342 green, and 1,524 blue galaxies. The color-magnitude cut used to
separate blue and red galaxies in \citet{Willmer06} for the DEEP2 sample is very
similar to the cut used here. We do not allow the definition of the center of
the green valley to evolve with redshift; however if a redshift evolution of
$\sim0.1$ per unit redshift \citep{Blanton06} is included in our color-magnitude
cut, none of our results or conclusions change.We do not create strictly
volume-limited samples, as we are not attempting to measure evolution in the
morphological parameters of green galaxies with redshift. We impose an absolute
magnitude limit of $M_B>-18$, to ensure that our colors samples (red, green and
blue) are of roughly similar depth. We show in Section~\ref{sec:casparam} that
the distribution of morphological parameters is not a strong function of
magnitude near the limit used here.

We test the robustness of our results to the color used to define the green
valley. In Figure~\ref{fig:colorcut} we show in the left panels samples defined
using (U-B) colors (in (U-B)-$M_B$ space in the top panel and in (NUV-R)-$M_B$
space in the bottom panel), along with samples defined using (NUV-R) colors in
the right panels (again with (U-B)-$M_B$ in the top panel and (NUV-R)-$M_B$ in
the bottom panel). To define the green valley in (NUV-R)-$M_B$ space, we fit for
the $M_B$ magnitude slope evolution by fitting a Gaussian to the blue galaxies
in four bins in magnitude. The color-magnitude cut is then shifted to the center
between the red and blue peaks. The width of the green valley in (NUV-R) color
is defined as $\delta (NUV-R)<0.45$, which is chosen to ensure that a similar
fraction of galaxies in the full sample is defined to be green in either (U-B)
or (NUV-R) space.

We find that galaxies defined to be in the green valley in (NUV-R) color also
lie in or near the green valley as defined in (U-B) color, with some scatter.
Galaxies that are defined to be green in one color that are \emph{not} green in
the other color still lie very close to the defined green valley. We have
verified that performing all of the analyses in this paper with an
(NUV-R)-selected sample does not change any of our results.

We note that the definition of the green valley in (NUV-R) color fit for and
used here is bluer than the definition in \citet{Salim09}, who defined green
galaxies as having $3.5<(NUV-R)<4.5$. Here our final green galaxy sample
(defined in (U-B)-$M_B$ space) lies within $3.2<(NUV-R)<4.1$.

%%%%%%%%%%%%%%%%%%%%%%%%%%%%%%%%%%%%%%%%%%%%%%%%%%%%%%%%%%%%%%%%%%%%%%%%%%%%%%%%
\section{Measured Morphology Parameters} \label{sec:parameters}
%%%%%%%%%%%%%%%%%%%%%%%%%%%%%%%%%%%%%%%%%%%%%%%%%%%%%%%%%%%%%%%%%%%%%%%%%%%%%%%%
We derive quantitative morphological parameters for all galaxies in our sample
from the \hst/ACS imaging. To sample rest-frame $B$-band morphologies across the
redshift range $0.4<z<1.2$, we use observed $V$-band morphologies for galaxies
with $z<0.6$ and observed $I$-band morphologies for galaxies with $z\ge0.6$.
Morphological classifications are flux limited to $I_{AB}<25$. Following
\citet{Lotz04} and \citet{Lotz06}, we measure morphologies for those objects
that have large enough sizes (Petrosian radius $\rp\ge0.3''$) and high enough
signal-to-noise (mean S/N per galaxy pixel $> 4$ within the segmentation map) to
yield robust morphological parameter measurements. The <S/N> threshold depends
on the pixel scale, while the Petrosian radius threshold depends on both the
pixel scale and the point spread function; therefore these thresholds are
\hst/ACS specific. The Petrosian radius $\rp$ is defined as the semi-major axis
length at which the ratio of the surface brightness at $\rp$ to the mean surface
brightness within $\rp$ is equal to 0.2. $\rp$ is computed within an elliptical
aperture using the ellipticity computed from the SExtractor galaxy photometry
software \citep{Bertin96}. The flux and radius cuts corresponds to an average
surface brightness within the Petrosian radius of $\mu \sim 24.4$ AB magnitudes
per square-arcsec \citep{Lotz08a}. All of the quantitative morphological
measurements below use the same segmented maps created by \citet{Lotz04}. With
these cuts, our sample contains 2,437 galaxies with measured morphologies and
redshifts within $0.4<z<1.2$. We derive a variety of morphological parameters;
details of each are given below.

\begin{figure*}
  \epsscale{1.}
  \plotone{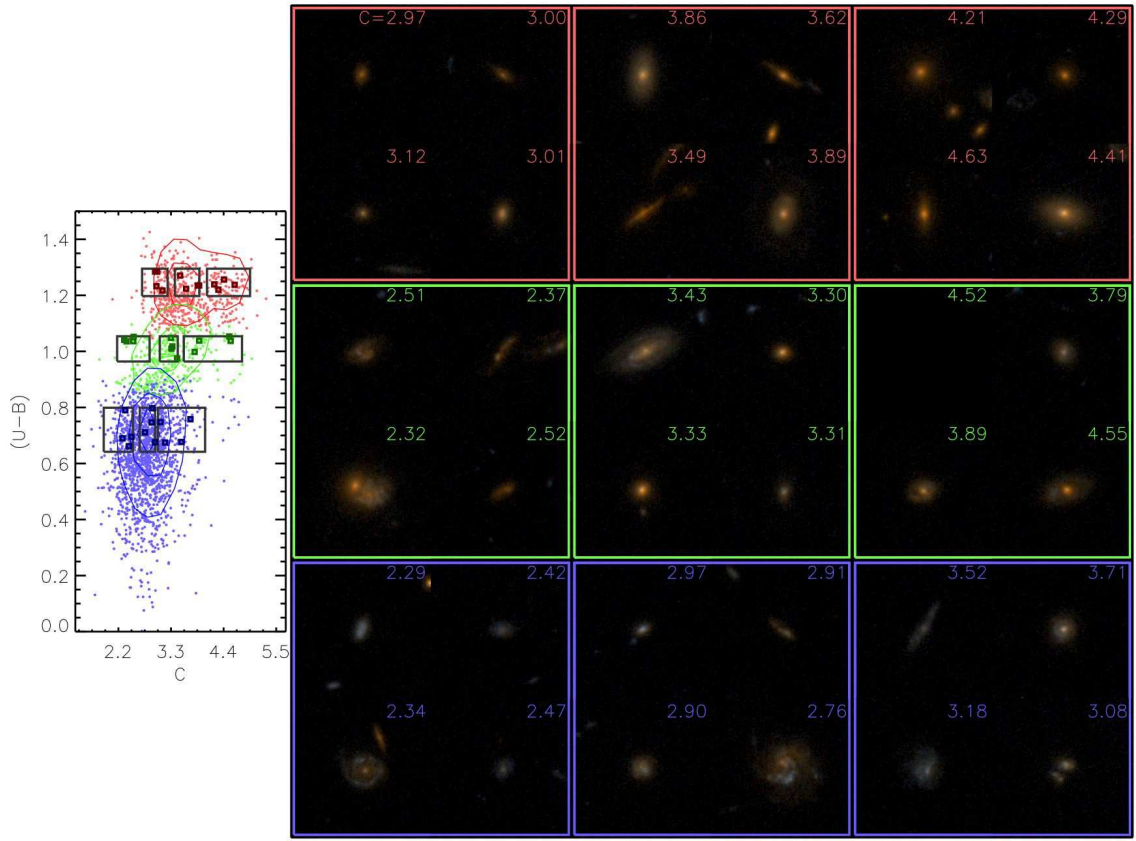}
  \caption{To help visualize the different morphology parameters that we include
  in this paper, we show here and in Figures 5 and 6 a random selection of
  \hst/ACS $V+I$ color image postage stamps spanning the full range of the
  concentration, asymmetry, and $\bt$ parameter for galaxies in each red, green
  and blue color space. To select the galaxies, each color space is divided into
  three areas, comprising the lower 20\% extreme of the cumulative relevant
  morphological distribution, middle 30\% centered at the median, and higher
  20\% extreme. The selection boxes are shown in the bivariate parameter and
  color plots to the left. Contours contain 30\%, 50\% and 80\% of the
  individual color-parameter distributions. From within each selection region we
  randomly draw four galaxies; postage stamps for each are shown on the right,
  with the value of the measured morphology parameter shown in the top right
  corner of each postage stamp for that galaxy. Postage stamps are $6\arcsec$
  across, corresponding to $\sim65$ kpc at $z=0.8$.}
  \label{fig:thirtysixc}
\end{figure*}

  \subsection{$\cas$ Parameters} We define concentration ($C$), asymmetry ($A$),
  and smoothness ($S$) following \citet{Conselice03}. The concentration
  parameter, which measures the central concentration of light in a galaxy
  \citep{Bershady00}, is defined as

  \begin{equation}
    C = 5\,{\rm log}\left(\frac{r_{80}}{r_{20}}\right)
  \end{equation}

  \noindent where $r_{80}$ and $r_{20}$ are the radii that contain 80\% and 20\%
  of the total light, respectively. Elliptical galaxies are generally the most
  concentrated with $C\sim4.5$, and concentration is found to decrease with
  later Hubble types to $C\sim2.5$ \citep{Conselice03}. In
  Figure~\ref{fig:thirtysixc}, we show \hst/ACS postage stamps (6$\arcsec$ on a
  side) of randomly-selected red, green and blue galaxies with relatively low,
  medium, and high concentration values within each color-selected sample. The
  left panel shows where the randomly-selected galaxies lie in
  color-concentration space, relative to the entire sample; the grey boxes show
  the regions used to select the galaxies with different concentration values.
  The concentration values for each galaxy shown are given in the upper right
  corner of each postage stamp.

  \begin{figure*}
    \epsscale{1.}
    \plotone{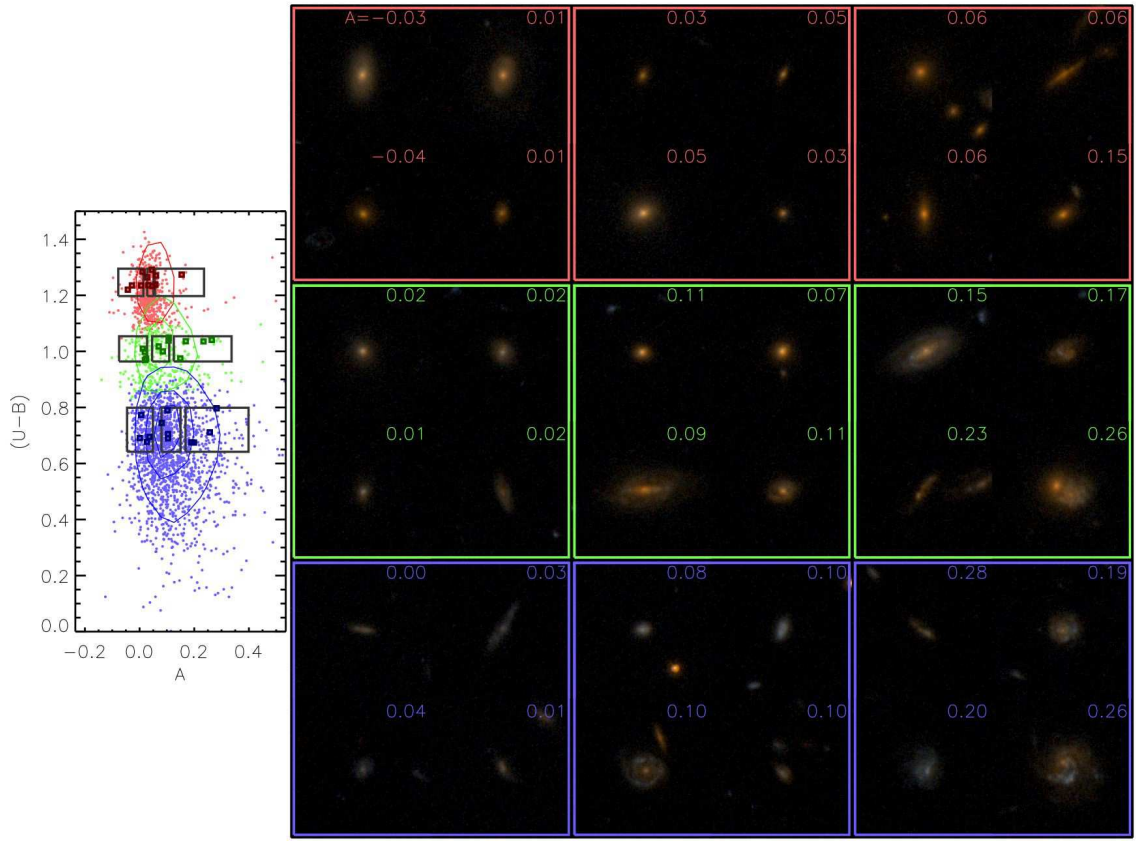}
    \caption{A random selection of $\hst$ postage stamps for galaxies
    spanning the measured asymmetry parameter (See Figure~\ref{fig:thirtysixc}
    for details).}
    \label{fig:thirtysixa}
  \end{figure*}

  Asymmetry is defined as the absolute value of the difference between the pixel
  intensities of a galaxy before ($I(i,j)$) and after a rotated by 180$\arcdeg$
  about its center ($I_{180}(i,j)$), normalized by the integrated intensity:

  \begin{equation}
    A = \frac{{\displaystyle \Sigma_{i,j}\|I(i,j)-I_{180}(i,j)\|}}
             {{\displaystyle \Sigma_{i,j}\|I(i,j)\|}} - B_{180}
  \end{equation}

  Here a correction is made to subtract off the averaged asymmetry of the
  background near the galaxy ($B_{180}$). \citep[see][for details]{Conselice00}.
  Asymmetry can be caused by spiral arms, dust lanes, mergers, or interactions,
  and is lowest with smooth elliptical light profiles and rises in star-forming
  and irregular galaxies, as well as on-going mergers \citep{Abraham94}. Major
  mergers typically have $A\ge0.35$, while spirals have $A\sim0.25$ and
  elliptical galaxies have $A\sim0.02$ \citep{Conselice03}. Similar to
  Figure~\ref{fig:thirtysixc}, in Figure~\ref{fig:thirtysixa} we show postage
  stamps for randomly-selected galaxies with different asymmetry values within
  the red, green, and blue galaxy populations.

  Smoothness is defined as the absolute value of the difference between the
  galaxy intensity pixel values ($I(i,j)$) and boxcar-smoothed intensity pixel
  values ($I_S(i,j)$) within $1.5\times\rp$ of the center of the galaxy:
  \begin{equation}
    S = \frac{{\displaystyle \Sigma_{i,j}\|I(i,j)-I_{S}(i,j)\|}}
             {{\displaystyle \Sigma_{i,j}\|I(i,j)\|}} - B_{S}.
  \end{equation}

  Here again the average smoothness of the background ($B_S$) is removed. Low S
  values correspond to smoother light profiles, while high S values correspond
  to less smooth light profiles. The smoothness parameter is sensitive to higher
  frequency clumps that are differentiated from the smoother parts of the
  galaxy. Smoothness correlates with patchiness, which can be due to recent star
  formation or compact star clusters \citep{Takamiya99}. An important issue with
  smoothness is its dependence on the apparent size of a galaxy and the
  smoothing length ($0.25\times\rp$), which can cause smaller galaxies to have
  smoothness values below the average smoothness of the background. This
  systematic issue with the smoothness parameter can limit its usefulness as a
  comparison parameter.

  \begin{figure*}
    \epsscale{1.}
    \plotone{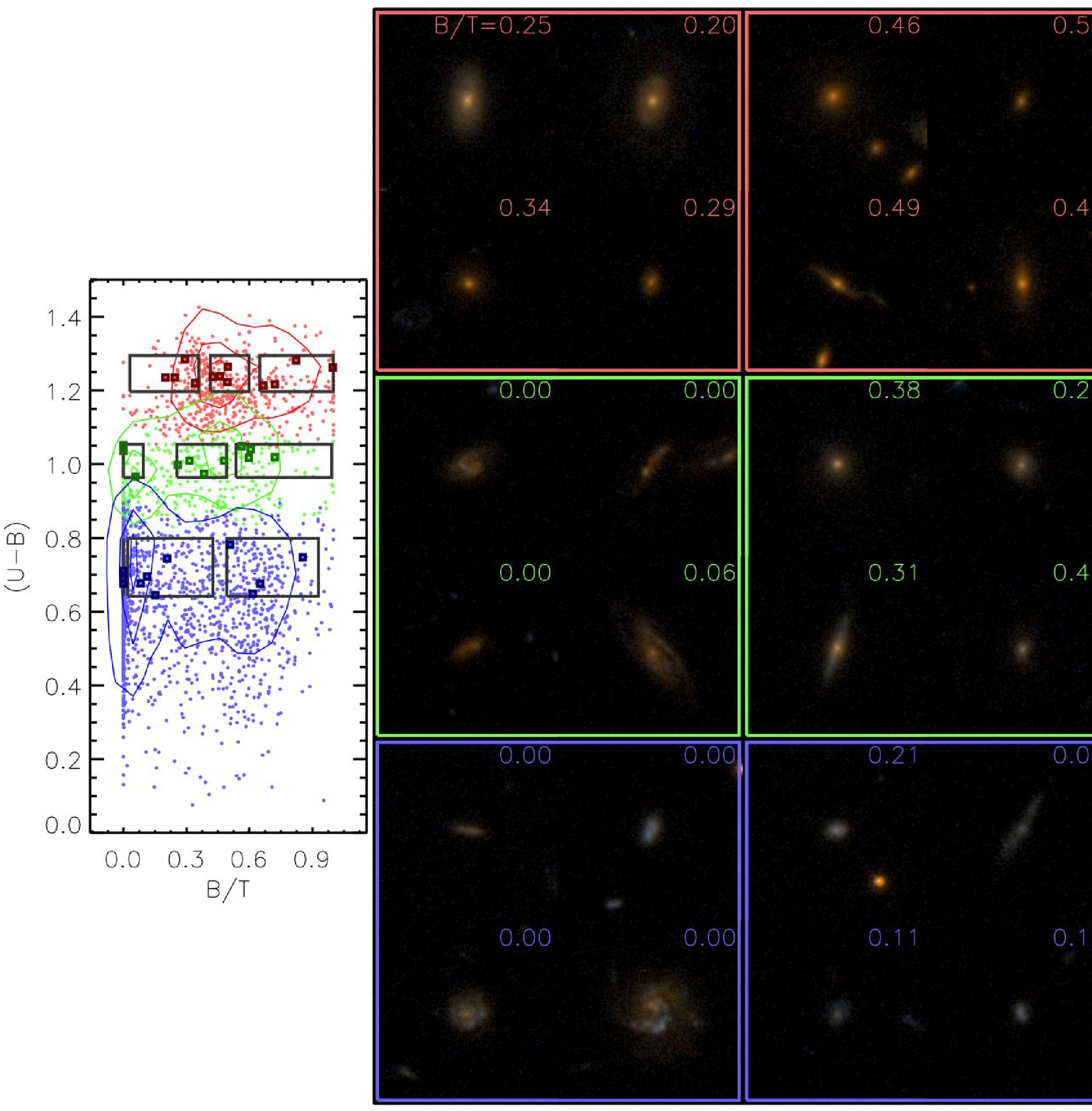}
    \caption{A random selection of $\hst$ postage stamps for galaxies spanning
    the measured $\bt$ parameter (See Figure~\ref{fig:thirtysixc} for details).}
    \label{fig:thirtysixbt}
  \end{figure*}

  \subsection{$\bt$: Bulge-to-Total Decomposition} Using two-dimensional bulge
  plus disk surface brightness profile models, we measure the bulge-to-total
  light fractions ($\bt$) using GIM2D \citep{Simard02}. The $\bt$ parameter
  provides a rough measure of how disk-dominated or bulge-dominated a galaxy is.
  The $\bt$ parameter is estimated using a S{\'e}rsic profile, with the
  S{\'e}rsic index ($n$) defined in \citet{Sersic68}, which controls the degree
  of curvature of the profile: a smaller $n$ reflects a less centrally
  concentrated profile with a steeper slope at large radii. Often standard $\bt$
  fits at low redshift use a classical de Vaucoulers profile with a S{\'e}rsic
  index of $n=4$ fit to the bulge component. Here we use $n=2$ to fit our higher
  redshift, younger galaxies, as many may not yet have formed a classical $n=4$
  bulge component. While using $n=2$ or $n=4$ can change the exact value of
  $\bt$ measured for a given galaxy, it does not change the trends or
  conclusions we find here. The only significant difference is that in
  Table~\ref{table:kstwo} we find a statistically different distribution between
  green and purple galaxies at the 5\% level (see Section~\ref{sec:kstests} for
  details). To be conservative, we adopt $n=2$ here.

  From the GIM2D V- and I-band best-fit model magnitudes of the entire galaxy
  ($I_{galaxy}$, $V_{galaxy}$) and bulge ($I_{bulge}$, $V_{bulge}$), the
  bulge-to-total fraction is defined as:

  $$
    \bt = \left\{
      \begin{array}{lll}
        10.0^{(V_{galaxy} - V_{bulge})/2.5} & {\rm for} & z<0.6 \\
        10.0^{(I_{galaxy} - I_{bulge})/2.5} & {\rm for} & z\ge0.6
      \end{array}
    \right.
  $$

  Figure~\ref{fig:thirtysixbt} shows postage stamps for red, green, and blue
  galaxies with a range of $\bt$ parameters, similar to
  Figures~\ref{fig:thirtysixc}-\ref{fig:thirtysixa}.

  \subsection{$G$: Gini Coefficient / $\mtwe$: Second-Order Moment of the 20\%
  of light.} In addition to the $\cas$ and $\bt$ morphological parameters, we
  also measure $G/\mtwe$ \citep{Lotz04}. The Gini coefficient ($G$) quantifies
  the distribution of light among the pixels in a galaxy and is defined as the
  absolute value of the difference between the integrated cumulative
  distribution of galaxy intensities and a uniform intensity distribution
  \citep{Abraham03}:

  \begin{equation}
    G = \frac{{\displaystyle \Sigma_i^n(2i-n-1)|X_i|}}
             {{\displaystyle |\bar{X}|n(n-1)}}
  \end{equation}

  \noindent where $n$ is the number of pixels, $X_i$ are the increasing ordered
  pixel flux values, and $\bar{X}$ is the mean pixel flux over the galaxy. $G$
  is close to unity if a single pixel contains all of the intensity and close to
  zero if the pixel values are uniform across the entire galaxy. $G$ correlates
  with concentration and surface brightness, but it is not sensitive to the
  \emph{location} of the brightest pixels. Therefore $G$ is high for galaxies
  with multiple bright nuclei as well as centrally concentrated spheroids.

  $\mtwe$ is the second order moment of the brightest 20\% of the galaxy light
  distribution, defined as the sum of the intensity of each pixel multiplied by
  the square of the distance from the center of the galaxy for the brightest
  20\% of the pixels in a galaxy. As such we can define $\mtwe$ as:

  \begin{eqnarray}
    M_{tot} &=& {\displaystyle \Sigma_i^n f_i [(x_i-x_c)^2 + (y_i-y_c)^2]} \\
    M_{20} &=& {\displaystyle {\rm log}\left(\frac{{\displaystyle \Sigma_i M_i}}
                                                   {{M_{tot}}}\right)
             {\rm,\quad with }\quad\Sigma_i f_i < 0.20 \ f_{tot}}
  \end{eqnarray}

  \noindent where $f_i$ are pixel fluxes and ($x_c$, $y_c$) is the location of
  the galaxy center that minimizes the total second order galaxy moment,
  $M_{tot}$. $\mtwe$ therefore traces the spatial extent of the brightest pixels
  and is anti-correlated with concentration.$\mtwe$ is typically $\sim-1.5$ for
  late-type galaxies and $\sim-2$ for early-type galaxies \citep{Lotz04}.

%%%%%%%%%%%%%%%%%%%%%%%%%%%%%%%%%%%%%%%%%%%%%%%%%%%%%%%%%%%%%%%%%%%%%%%%%%%%%%%%
\section{Green Galaxy Morphologies} \label{sec:results}
%%%%%%%%%%%%%%%%%%%%%%%%%%%%%%%%%%%%%%%%%%%%%%%%%%%%%%%%%%%%%%%%%%%%%%%%%%%%%%%%
We first study various individual and joint distributions of both the general
properties of green galaxies and their morphological properties, and compare
them to the distributions of red and blue galaxies. The goal of this section is
to understand the properties of galaxies in the green valley and how they differ
from both red and blue galaxies. 

  % Galaxy Properties: ---------------------------------------------------------
  \begin{figure*}
    \epsscale{0.9}
    \plotone{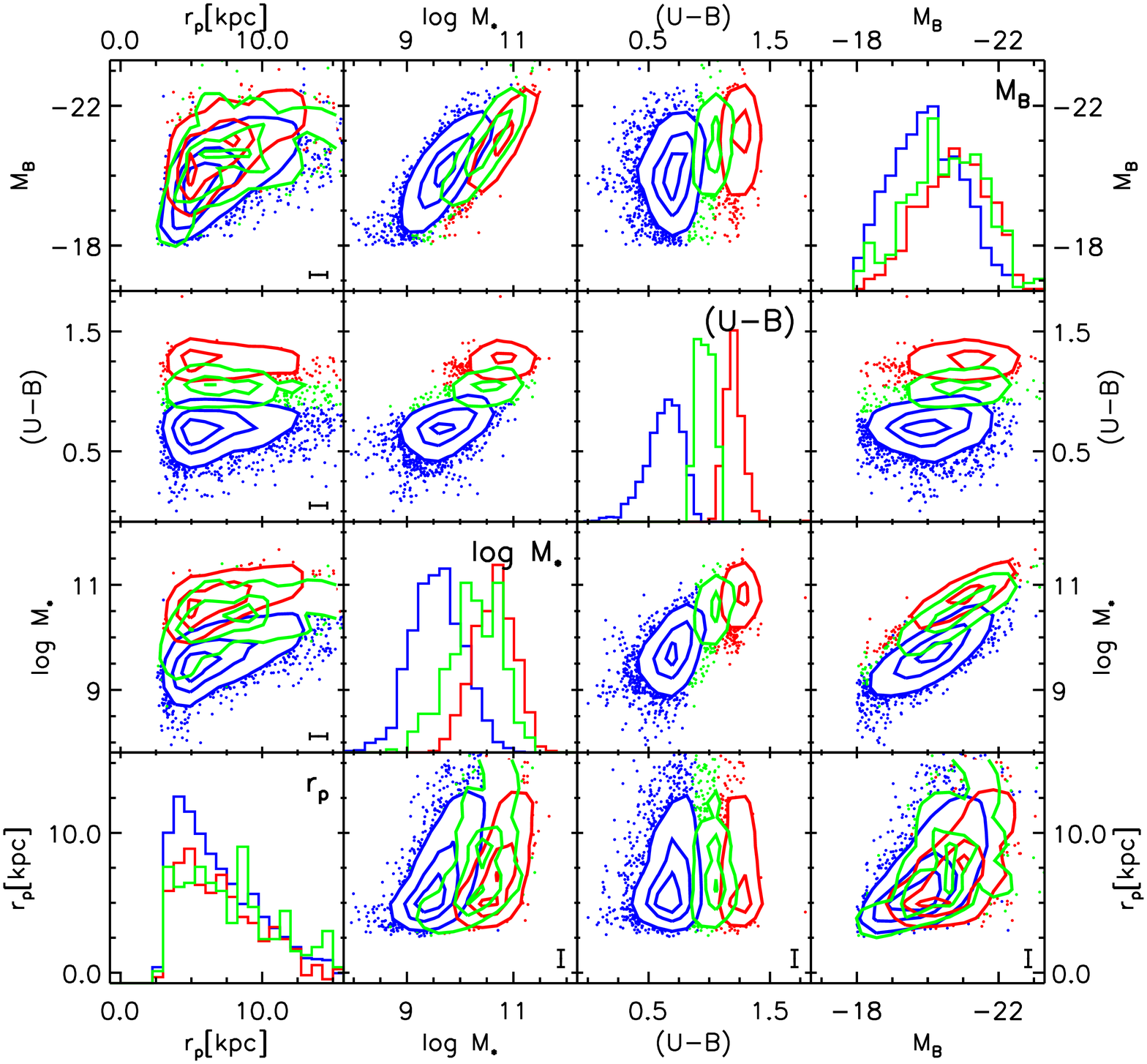}
    \caption{General properties of the AEGIS galaxy sample used here in the
    joint space of absolute $B$-band magnitude $M_B$, rest-frame (U-B) color,
    stellar mass, and Petrosian radius ($\rp$ in units of kpc). Along the
    diagonal are individual parameter histograms for each color-selected sample,
    where we have weighted the red and blue galaxy histograms by a half and a
    quarter, respectively, to facilitate comparison with the green galaxies. Off
    diagonal panels show bivariate joint distributions, with contours containing
    30\%, 50\%, and 80\% of the sample; galaxies outside the 80\% contour are
    shown as individual points. Contours have been smoothed using a Gaussian
    1/20$^{\rm th}$ of the size of the panel. The innermost 30\% contour is not
    plotted if it is smaller than the grid used to calculate the contours.
    1$\sigma$ error bars at the median S/N pixel$^{-1}$ of the sample, estimated
    from \citet{Lotz06}, are shown in the lower right corner of each space where
    available.}
    \label{fig:properties}
  \end{figure*}

  \subsection{Galaxy Properties}\label{sec:properties}
  Figure~\ref{fig:properties} shows the general properties of red, green, and
  blue galaxies in the joint spaces of absolute $B$-band magnitude $M_B$,
  rest-frame (U-B) color, stellar mass, and Petrosian radius ($\rp$), which we
  use to estimate the size of the galaxy. Along the diagonal in this figure are
  individual parameter histograms for each color-selected sample, while
  off-diagonal panels show bivariate joint distributions.

  The green galaxy sample has both an absolute magnitude $M_B$ and stellar mass
  distribution that is intermediate between the red and blue galaxy samples.
  This is not surprising given the dependence of stellar mass on both color and
  magnitude. The median stellar mass of the red, green, and blue galaxy samples
  are fairly different, at 10.7, 10.3, and 9.6 log $M_*/M_{\sun}$, respectively.
  In addition, there is a well-known strong correlation between size, magnitude,
  and stellar mass, in that brighter and/or more massive galaxies have larger
  sizes \citep{Blanton09}. However there is not a strong correlation between
  (U-B) color and Petrosian radius, except within the blue population. The green
  galaxy population on the whole has similar size and magnitude distributions to
  red galaxies. We note that the green galaxy sample has a tail to larger
  Petrosian radii than either the blue or red galaxy samples. Interestingly,
  \citet{Salim10} find using \hst/ACS imaging that UV-excess, early-type
  galaxies that fall mainly in the green valley at low redshift are, on average,
  larger than either red or blue galaxies. We discuss the statistical
  significance of this result in Section~\ref{sec:kstests}. In
  Section~\ref{sec:purple}, we will address possible stellar mass-dependent
  morphological differences between these samples by comparing samples with
  similar stellar mass distributions.

  \subsection{$\cas$ and $\bt$ Parameter Results}\label{sec:casparam} 
\begin{figure*}
    \epsscale{1.2}
    \plotone{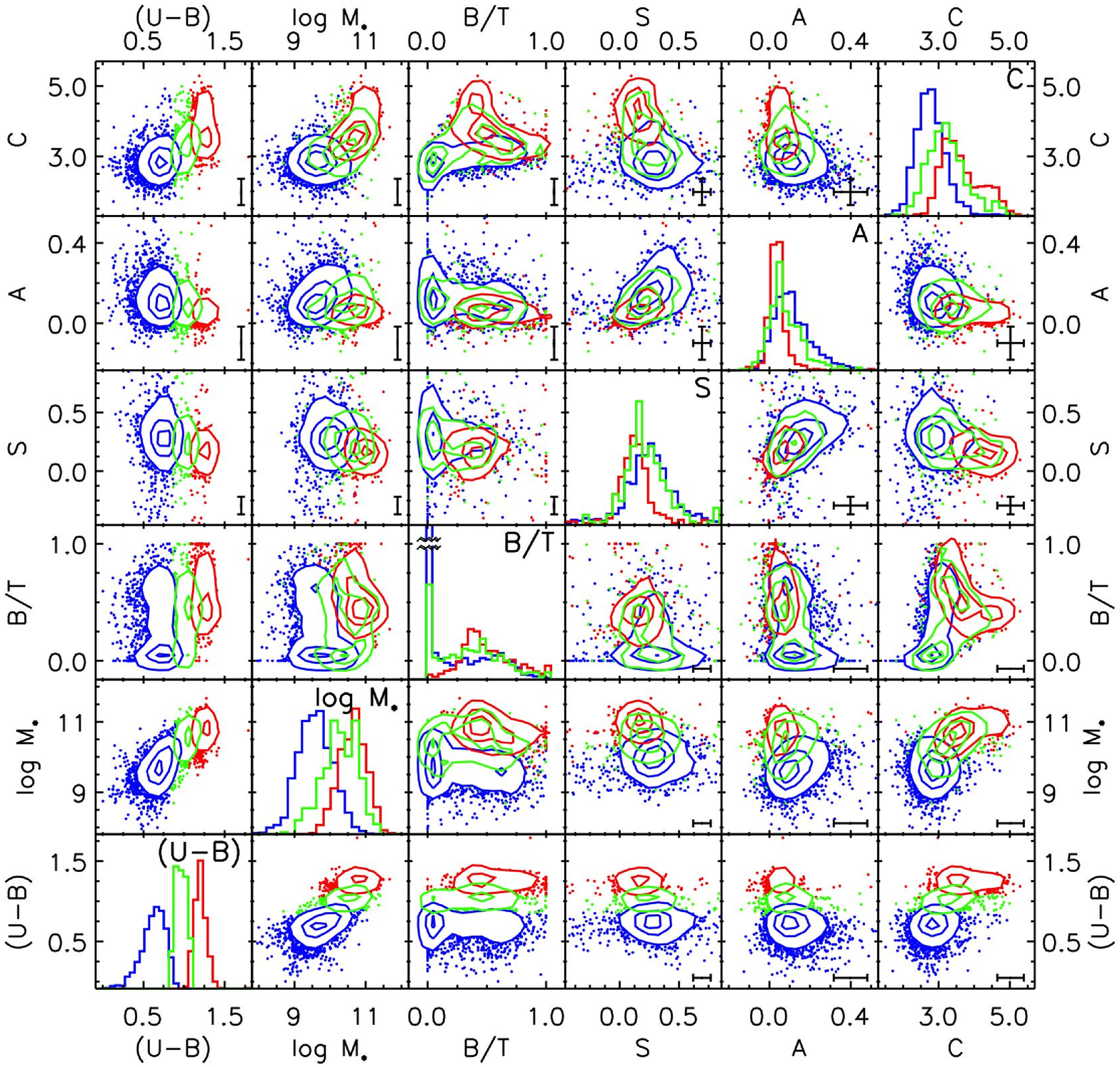}
    \caption{Morphological properties of red, green, and blue galaxies at
    $0.4<z<1.2$ in AEGIS. Shown are joint distributions of galaxy rest-frame
    (U-B) color, stellar mass, concentration ($C$), asymmetry ($A$), smoothness
    ($S$), and bulge-to-total light fraction ($\bt$). Along the diagonal are
    distributions for each individual parameter, where the red, and blue
    distributions have been scaled down by a factor of two, and four,
    respectively. Due to the large fraction of bulge-less ($\bt=0$) blue
    galaxies, we limit the range of $\bt$ to be 50\% of the full range. The
    bivariate parameter distributions have contours plotted at 30\%, 50\%, and
    80\% of the total sample. The innermost 30\% contour is not plotted
    if it is smaller than the grid used to calculate the contours. 1$\sigma$
    error bars at the median S/N pixel$^{-1}$ of the sample, estimated from
    \citet{Lotz06}, are shown in the lower right corner of each space where
    available.}
    \label{fig:casparam}
  \end{figure*}

We next
  study the morphological distributions of green galaxies in $\cas$ and $\bt$
  space and compare them with samples of red and blue galaxies.
  Figure~\ref{fig:casparam} shows the distribution of red, green, and blue
  galaxies in terms of $\cas$, $\bt$, stellar mass, and (U-B) color with error
  bars, contours, and histograms similar to Figure~\ref{fig:properties}. Error
  bars on morphological parameters are estimated from \citet{Lotz06}. We begin
  by investigating the individual parameter space distributions along the
  diagonal in Figure~\ref{fig:casparam}.

  % CAS + B/T + Stellar mass + Color:  -----------------------------------------

  In terms of the concentration parameter, blue galaxies have $C\sim2-3.5$,
  while red galaxies have $C\sim3-5$. Green galaxies have an intermediate
  distribution, with $C\sim2.5-4$. Unlike red galaxies, the green galaxy
  population does \emph{not} contain a large fraction of galaxies with
  particularly high concentration values; 10\% of green galaxies have $C>4$
  compared to 26\% of red galaxies.

 	The fraction of galaxies with high asymmetry is dependent on color. Blue
  galaxies have larger asymmetry values than red galaxies on the whole. Using
  the threshold of $A>0.35$ that is typically used to define mergers, for
  example, we find that 3\% of blue galaxies and 0.4\% of red galaxies have
  $A>0.35$. Using a threshold of $A>0.2$, which clearly separates the tail of
  the distribution, we find that 17\% of blue galaxies and 1.2\% of red galaxies
  have $A>0.2$. The green galaxy population has a similar asymmetry distribution
  as the blue galaxy population, though fewer green galaxies have large $A$
  values (1.2\% have $A>0.35$ and 8\% have $A>0.2$). Using either high $A$
  threshold, we find a lower percentage of highly asymmetric green galaxies
  compared to blue galaxies. This implies that while green galaxies have
  similar amounts of asymmetry as blue galaxies due to star formation knots,
  dust lanes, and/or spiral arms, they do not have as high a fraction of major
  mergers as the blue galaxy population (see Section~\ref{sec:ginitype} for a
  discussion of the merger fraction in each population using $G$-$\mtwe$).

  To determine robustly the smoothness parameter of a galaxy one requires a
  larger threshold in spatial extent than for measuring concentration or
  asymmetry. The threshold required for smoothness is $\rp\ge0.6\arcsec$,
  compared to $\rp\ge0.3\arcsec$ required for concentration and asymmetry.
  Therefore smoothness can be measured for only roughly half of the galaxies in
  our sample. We find that red galaxies have lower $S$ values than blue
  galaxies, on the whole, which reflects the fact that red galaxies generally
  have smoother light profiles. The green galaxy population has a somewhat
  similar $S$ distribution as the blue galaxy population, though with more
  galaxies having lower $S$ values. The median $S$ values of red, green, and
  blue samples are 0.12, 0.19, and 0.24, respectively.

  Continuing down the diagonal of Figure~\ref{fig:casparam} to the $\bt$
  parameter, we find that while blue galaxies have a range of $\bt$ values from
  0 to $\sim0.8$, a large fraction (30\%) have $\bt=0$, implying a disk-only
  galaxy with no bulge component. In Figure~\ref{fig:casparam}, the smallest
  $\bt$ bin contains that 44\% of blue galaxies with $0\le\bt<0.05$. By-eye
  inspection of the \hst/ACS images of these bulge-less galaxies confirms that
  these are disk-only systems. Due to this enhancement at $\bt=0$ for the blue
  population, we limit the y-axis on the histogram plot of $\bt$ to 50\% of the
  peak number of blue galaxies at $\bt=0$ to clearly show the full distribution
  of each color sample. Red galaxies have larger $\bt$ fractions than blue
  galaxies, on the whole, with a distribution centered at $\bt\sim0.4$, and very
  few bulge-less galaxies (0.9\%, 1.5\% with $0\le\bt<0.05$). Green galaxies
  have a $\bt$ distribution that is similar to blue galaxies, though with a
  smaller fraction (12\%, 21\% with $0\le\bt<0.05$) of bulge-less galaxies.
  Figure~\ref{fig:thirtysixbt} includes \hst/ACS postage stamps of four
  randomly-selected bulge-less green galaxies in the central left sub-panel.

  These bulge-less green galaxies are particularly interesting as the creation
  mechanism for these galaxies is unclear. We have visually inspected the
  \hst/ACS images of these sources to check that they indeed have no apparent
  bulge component and are disk-only systems. We further investigate whether this
  population is dominated by dusty star-forming galaxies. In the (NUV-R) versus
  SSFR space shown in Figure~\ref{fig:nuvr}, these bulge-less green galaxies
  (outlined by black diamonds) span the entire green galaxy population. We find
  that there is not a statistically significant difference in the dusty fraction
  of green galaxies with $\bt=0$ ($27\pm5\%$) compared to the entire green
  galaxy population ($21\pm3\%$). Therefore the bulk of this population should
  be transition objects from the blue cloud to the red sequence. As these
  galaxies do not have a bulge, it is unlikely that they have central AGNs which
  could lead to strong feedback. It is also unlikely that they have had many
  major mergers, which presumably would have led to the creation of a bulge
  component. The existence of these bulge-less green galaxies therefore places
  strong constraints on the quenching mechanism(s) at work in this population.

  In addition to the individual parameter distributions shown in
  Figure~\ref{fig:casparam}, we can study the bivariate distributions between
  various morphological parameters. In $C$-$A$, $A$-$S$, and $C$-$S$ space, in
  general green galaxies are an intermediate population between red and blue
  galaxies. The greatest differences between green galaxies compared to red and
  blue galaxies are seen in concentration and to a lesser extent in asymmetry.
  In Section~\ref{sec:kstests} we discuss the statistical significance of these
  differences since they may provide insight into the underlying physical
  mechanism(s) that shut off star formation in galaxies.

  % Concentration Plot: --------------------------------------------------------
  \begin{figure}
    \epsscale{1.2}
    \plotone{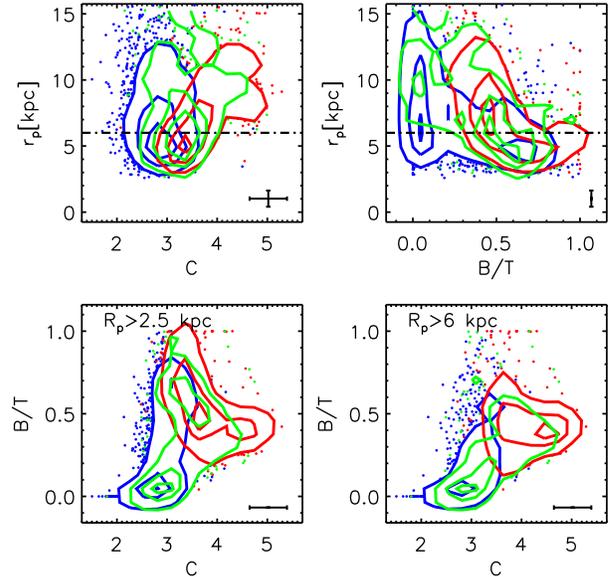}
    \caption{An investigation of how the measured $C$ and $\bt$ parameters
    depend on galaxy size. The upper panels show the measured Petrosian radius
    ($\rp$) as a function of $C$ and $\bt$ for our red, green, and blue galaxy
    samples. The dashed line at $\rp=6$ kpc shows the threshold used in the
    lower right panel. Contours and error bars are similar to
    Figure~\ref{fig:casparam}. The lower left panel shows $\bt$ versus $C$ for
    all galaxies with $\rp>2.5$ kpc. Both green and red galaxies show an odd
    trend of low $C$ and high $\bt$. In the lower right panel only galaxies with
    \rp$>6$ kpc are included and this trend disappears. }
    \label{fig:concentration}
  \end{figure}

  There are interesting differences between the galaxy populations in the
  $\bt$-$C$ parameter space. Within the blue galaxy population there is very
  little correlation between $\bt$ and $C$, except that the bulge-less ($\bt=0$)
  galaxies have the lowest concentration. However, within the red galaxy
  population there is an odd correlation in that the galaxies with the highest
  $\bt$ values have the lowest concentration values. This is likely not real but
  is due to these red galaxies with low $C$ and high $\bt$ values being very
  compact, such that $C$ is not well-measured. Comparing $C$ with $\rp$ for each
  color sample (see Figure~\ref{fig:concentration}) we find that within the blue
  and green populations there is no correlation, but within the red population
  the galaxies with $\rp<6$ kpc have small measured concentrations, with $C<4$.
  These galaxies also have the highest $\bt$ values, with $\bt>0.7$. Keeping
  only galaxies with $\rp>6$ kpc, we find that very few sources have $\bt>0.6$
  and within the red population there is no longer a correlation between $\bt$
  and $C$. 

  Green galaxies in the $\bt$-$C$ space have a distribution that is more similar
  to blue galaxies than red galaxies, though there is a tail with high
  concentration, as seen in the red galaxy sample. For a given $\bt$ value,
  green galaxies have higher concentrations than blue galaxies. This trend holds
  in the sample of galaxies with $\rp>6$ kpc. We discuss the implications of
  this in Section~\ref{sec:discussion}.

  Finally, in the $M_{*}$-$C$ parameter space, we find that the red and blue
  galaxy populations lie in distinct regions of this space. The green galaxy
  population appears to span the red and blue distributions, possibly as a
  transition population moving from low stellar mass and low concentration (like
  the blue galaxies) to higher stellar mass and higher concentration (like the
  red galaxies).

  % Conditional Distribution Plots: ----------------------------------------------
  \begin{figure}
    \epsscale{1.2}
    \plotone{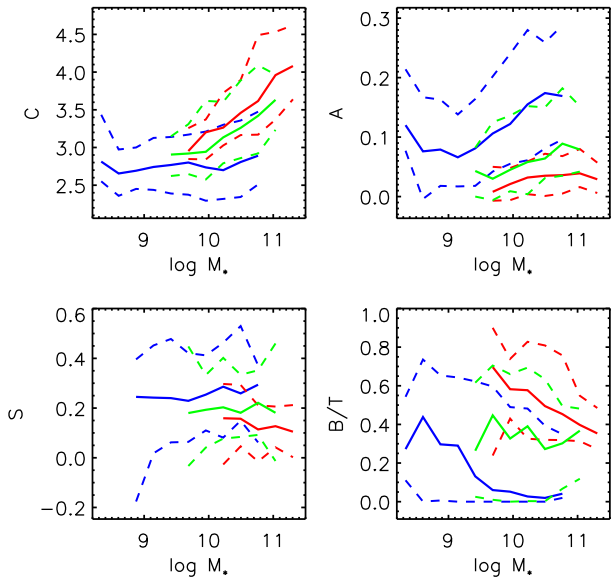}
    \caption{Conditional distribution plots which show the median and 68\% range
    for the red, green and blue galaxies. The four subplot show concentration,
    asymmetry, smoothness, and bulge to total fraction values at a given mass.
    For a fixed stellar mass Green galaxies have higher concentration, and lower
    asymmetry values as compared to the blue galaxies.}
    \label{fig:conditional}
  \end{figure}

  In Figure~\ref{fig:conditional} we highlight these differences by showing the
  $C$, $A$, $S$, and $\bt$ morphological distributions (mean and 68\% range) of
  each of the red, green, and blue galaxy samples as a function of stellar mass.
  This figure shows that at a given stellar mass green galaxies have larger C
  values, lower A values, and lower S values than blue galaxies. The B/T values
  may be higher than blue galaxies at a given stellar mass, but the noise in the
  mean and the similar width of the distributions make it difficult to draw a
  strong conclusion.

  \subsection{Rough Morphological types using $G$/$\mtwe$}\label{sec:ginitype}
  From the work of \citet{Lotz04} and \citet{Lotz08a}, galaxies can be
  classified into rough morphological types: early, late or merger, depending on
  their location in $G$-$\mtwe$ space (see Figure~\ref{fig:ginitype} for
  the cuts shown in the space). Following \citet{Lotz08a}, we define

  \begin{eqnarray*}
    {\rm Mergers}:         G &>& -0.14 \mtwe+0.33,\\
    {\rm Early (E/S0/Sa)}: G &\le& -0.14 \mtwe+0.33, \, {\rm and} \\
                           G &>& 0.14 \mtwe + 0.80,\\
    {\rm Late (Sb-Ir)}:    G &\le& -0.14 \mtwe+0.33, \, {\rm and}\\
                           G &\le& 0.14 \mtwe + 0.80,
  \end{eqnarray*}

  \noindent in the $G$/$\mtwe$ plane, which are based on visually determined
  morphologies in the EGS \citep{Lotz08a}. Note that mergers in this plane
  include both major and minor mergers, whereas high asymmetry ($A>0.35$) is
  only sensitive to major mergers \citep{Lotz04}.

  % RGB Type Table -------------------------------------------------------------
  \begin{deluxetable}{rrrr}  %Generated by textable.pro%
    \tablecolumns{4}
    \tablewidth{0pt}
    \tablecaption{$\rgb$ $G$/$\mtwe$ Morphological Types\label{table:ginirgb}\tablenotemark{a}}
    %SNIP%
    \tablehead{ & \colhead{Red} & \colhead{Green} &    \colhead{Blue}  }
    \startdata
	      Mergers & 12 $\pm$ 2 \% &   14 $\pm$ 2 \% &     19 $\pm$ 1 \% \\
	   Early Type & 66 $\pm$ 5 \% &   35 $\pm$ 4 \% &      8 $\pm$ 1 \% \\
	    Late Type & 22 $\pm$ 2 \% &   51 $\pm$ 5 \% &     73 $\pm$ 3 \% \\
	  \enddata
    %SNIP%
    \tablenotetext{a}{Percentage of each galaxy sample with $0.4<z<1.2$
    classified into morphological types using $G$/$\mtwe$. Uncertainties include
    both Poisson error and uncertainties in the measured values of $G$/$\mtwe$.
    These fractions are measured over a wide redshift range in samples that are
    not volume-limited and thus should not be taken as the global fractions for
    red, green and blue galaxies. Please see Table 3 in \cite{Lotz08a} for a the
    redshift-evolving fraction. These results here are shown to emphasize the
    relative fractions in our red, green, and blue galaxy samples.}
  \end{deluxetable}

  Table~\ref{table:ginirgb} contains the fraction of galaxies in the red, green,
  and blue galaxy populations used here that is classified as early
  type(E/S0/Sa), late type(Sb-dI), or merger. We derive errors on these
  fractions by performing Monte Carlo tests using the median errors estimated
  for $G$ and $\mtwe$ from the median S/N pixel$^{-1}$ \citep{Lotz06}. We also
  include in quadrature Poisson errors. Note that these fractions should not be
  interpreted as the fraction for all galaxies at these redshifts, as our
  samples are not strictly volume-limited and they cover a wide redshift range.
  We refer the reader to Table~3 in \cite{Lotz08a} for the redshift-evolving
  morphological fraction of volume-limited color-defined galaxy samples at these
  redshifts. Here we use the derived fractions to compare relative trends
  between the different color-selected galaxy samples.

  From Table~\ref{table:ginirgb} we find that the green galaxy population is
  intermediate between the red and blue galaxy populations in terms of each of
  the three morphological types. Compared to the green galaxy population, the
  blue population has a statistically higher merger fraction ($2\sigma$), more
  late type galaxies ($4\sigma$), and fewer early type galaxies
  ($7\sigma$), while the red population has a lower merger fraction
  ($1\sigma$), fewer late type galaxies ($5\sigma$), and more early type
  galaxies ($5\sigma$). Comparing our results with the fractions found by
  \citep{Lotz08a}, we find the same general trends, though we have somewhat
  higher merger fractions for all three galaxy samples.

  \subsection{Rotated Gini and $\mtwe$ Parameters}\label{sec:rotatedGini} 
\begin{figure*}
    \epsscale{1.1}
    \plotone{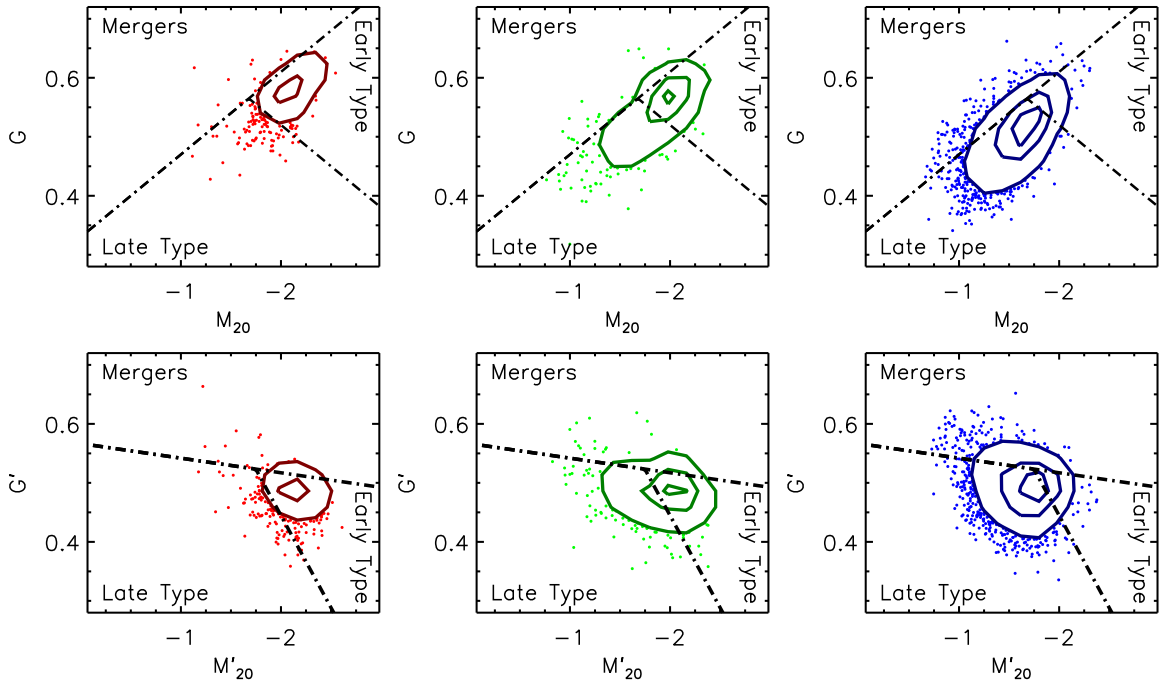}
    \caption{Using the cuts defined in \citet{Lotz04}, we divide the $G$/$\mtwe$
    space (top row) into areas with galaxies that are early-type (E/S0/Sa),
    late-type (Sb/Sc/Ir), or mergers and plot the locations of red, green, and
    blue galaxies. Contours and error bars are similar to
    Figure~\ref{fig:casparam}. The bottom row shows the same galaxy samples in a
    rotation version of this space in which the locus of galaxies lies along the
    $\gtheta$ axis (see text in Section~\ref{sec:rotatedGini} for details).
    Using this rotated definition, the $\mtheta$ axis relates more directly to
    the negative of the S{\'e}rsic index and $\gtheta$ correlates with merger
    identification. Due to the tight grouping of the red galaxy population in
    these spaces, the innermost 30\% contour is not shown.}
    \label{fig:ginitype}
  \end{figure*}

We
  also analyze our samples in the $\gtheta/\mtheta$ parameter spaces, which are
  plane rotated versions of $G$/$\mtwe$. These versions are created by rotating
  $G$/$\mtwe$ such that the locus of galaxies, spanning from late to early
  types, now lie along the $\mtheta$ axis. The locus is roughly parallel to the
  merger definition line and is perpendicular to the late and early type
  definition line. In Figure~\ref{fig:ginitype}, we plot red, green, and blue
  galaxies in $G$/$\mtwe$ bivariate space in the top row, along with the
  corresponding rotated $\gtheta/\mtheta$ space in the bottom row. 30\%, 50\%
  and 80\% contours are plotted, along with the morphological classification
  lines, for both the standard and rotated spaces. This rotation will not effect
  the rough morphological types found in Table~\ref{table:ginirgb}, but it
  breaks the degeneracy between the $G$ and $\mtwe$ parameters, such that the
  $\gtheta$ parameter reflects merger activity, while $\mtheta$ correlates with
  S{\'e}rsic index. The rotated variables are defined as

  \begin{equation*}
    \begin{bmatrix} \mtheta \\ \gtheta\end{bmatrix} = 
    \begin{bmatrix} {\rm cos }(\theta) & -{\rm sin}(\theta)\\
                    {\rm sin}(\theta) & {\rm cos }(\theta)
    \end{bmatrix} \times \begin{bmatrix}\mtwe\\G\end{bmatrix}
  \end{equation*}

  \noindent with $\theta = 0.164$ radians fit from the locus of galaxies in the
  $G$/$\mtwe$ plane.

  % Gini/M20 Type: -------------------------------------------------------------

\begin{figure*}
    \epsscale{1}
    \plotone{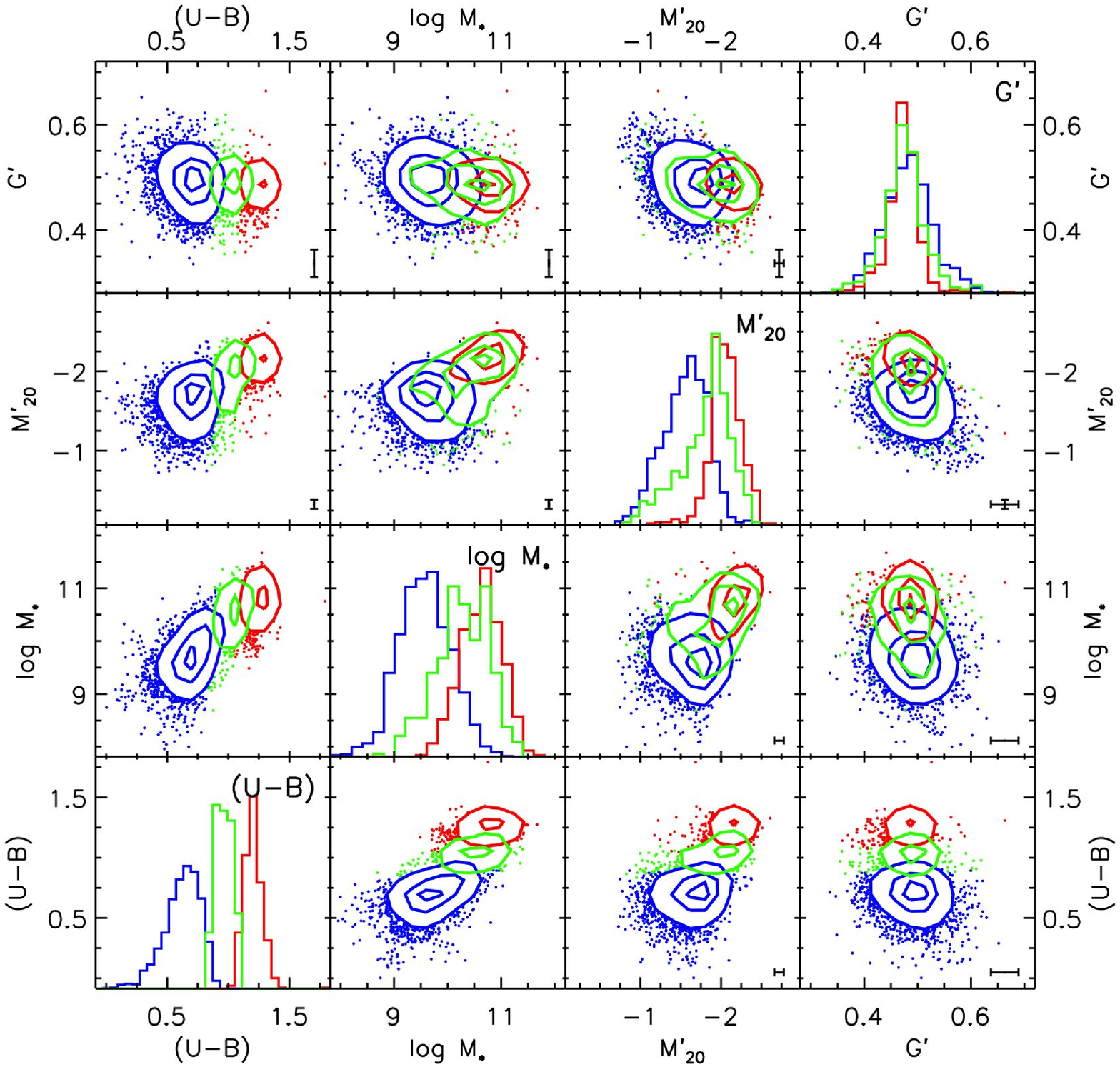}
    \caption{Measured $\gtheta/\mtheta$, stellar mass, and (U-B) color joint
    parameter spaces. Contours, histograms, and error bars are similar to Figure
    ~\ref{fig:casparam}.}
    \label{fig:giniparam}
  \end{figure*}

  Similar to Figure~\ref{fig:casparam}, in Figure~\ref{fig:giniparam} we
  consider the rotated $\gtheta$/$\mtheta$ space and joint distributions with
  both stellar mass and (U-B) color. Contours, histograms, and error bars are
  similar to Figure~\ref{fig:casparam}.

  % GINI/M20 + Rotated + Stellar mass + Color ----------------------------------

  In the one-dimensional $\gtheta$ space (upper right panel of
  Figure~\ref{fig:giniparam}), the blue population contains a wider distribution
  that extends to both larger and smaller values of $\gtheta$ compared to the
  red galaxy population. As higher $\gtheta$ values correlate with merger
  activity (reflecting both major and minor mergers), this implies a larger
  merger fraction for blue galaxies compared to red galaxies. Small values of
  $\gtheta$ can result from very dusty galaxies. The green galaxy population has
  an intermediate distribution in $\gtheta$ between the blue and red galaxy
  distributions. In particular, the green galaxy population has a lower fraction
  of objects with high $\gtheta$ values compared to blue galaxies and a higher
  fraction than red galaxies. This implies a merger fraction that is lower than
  what is found for the blue galaxy population but higher than the red galaxy
  population.

  Similar to the concentration parameter distribution, the green galaxy
  population is intermediate between the red and blue galaxies in the $\mtheta$
  parameter space, where red galaxies have a higher mean value of $\mtheta$ than
  blue galaxies. There is a sharp distinction between the red and blue galaxy
  populations in $\mtheta$ which is more pronounced in $\mtheta$ compared to
  $\gtheta$. While the green galaxy distribution spans the majority of both the
  red and blue populations, it does not include the highest values of $\mtheta$
  seen in the red galaxy distribution or the lowest values seen in the blue
  galaxy distribution.

  In the joint $\mtheta$-$\gtheta$ space, differences between the red, green,
  and blue populations are more distinct in $\mtheta$ than $\gtheta$. While
  there is very little correlation seen between stellar mass and $\gtheta$,
  there is a strong correlation between stellar mass and $\mtheta$. As with
  stellar mass and concentration, here again the red and blue galaxy populations
  lie in distinct areas of $M_{*}$-$\mtheta$ space. The green galaxy population
  appears to be intermediate, although their distribution is more similar to red
  galaxies than blue galaxies. For a given stellar mass, green galaxies have
  lower $\mtheta$ values than red galaxies, and the green galaxy population on
  the whole has many fewer galaxies with low $\mtheta$, as seen for the blue
  galaxy population.

  \subsection{Environmental Dependence of the Color-Morphology Relation} We
  further investigate the joint environment-stellar mass-morphology distribution
  of red, green, and blue galaxies by comparing the morphological distributions
  as a function of stellar mass of each galaxy color sample in different
  environments. Using the projected third-nearest-neighbor density catalogs of
  \citet{Cooper06}, we estimate the local environment or over density of each
  red, green, and blue galaxy in our sample. We find no significant difference
  in the morphology-stellar mass distribution of galaxies of a given color when
  split into roughly equal-sized populations in over- or under-dense regions. To
  compare the morphology-stellar mass distributions of galaxies in the most
  extreme environments within our sample, we select galaxies such that they have
  $|{\rm log}_{10}(1+\delta_3)|>0.25$, and do not find any significant
  differences in the C, A or $\bt$ parameters. We additionally use the DEEP2
  group catalogs of \citet{Gerke07} to identify galaxies likely to be in groups,
  with velocity dispersions $\sigma_v \ge 150$ km s$^{-1}$ or in the field. We
  compare the morphology-stellar mass distributions of red, green, and blue
  galaxies in groups versus the field and find no significant differences,
  though we note that the errors are somewhat large due to small numbers of
  galaxies in each color-stellar mass-morphology-environment bin.

%%%%%%%%%%%%%%%%%%%%%%%%%%%%%%%%%%%%%%%%%%%%%%%%%%%%%%%%%%%%%%%%%%%%%%%%%%%%%%%%
\section{Green versus Purple Galaxy Comparison} \label{sec:purpleresults}
%%%%%%%%%%%%%%%%%%%%%%%%%%%%%%%%%%%%%%%%%%%%%%%%%%%%%%%%%%%%%%%%%%%%%%%%%%%%%%%%
The second goal of this paper, beyond measuring the morphological distributions
of green valley galaxies, is to quantitatively determine whether the green
galaxy population can be explained from a morphological point of view as a
simple mix of blue and red galaxies, or whether green galaxies have a distinct
morphological make-up that defines them as a separate, distinct population. To
undertake this second goal, we wish to compare the morphological distribution of
green galaxies with the distribution of red and blue galaxies taken as a whole.
However, we can not simply take the union of the red and blue galaxy samples
used here to compare with the green galaxy sample, as the morphologies of the
red and blue galaxy populations differ significantly. As a result, the
morphological distribution of their union will depend on the ratio of red to
blue galaxies in the combined sample. We therefore create ``purple'' galaxy
comparison samples that have the same ratio of red to blue galaxies as within
the green population itself. We also create comparison samples that have had
dusty galaxies removed and have matched stellar mass distributions, to limit
possible biases in our statistical tests.

We found in Section~\ref{sec:properties}, when comparing red, green, and blue
galaxies in various morphological parameter spaces that the green galaxy
population appears to be intermediate between the red and blue galaxy
populations. Using the purple comparison samples, we can test statistically if
green galaxies are in a transition phase, moving from the blue cloud to the red
sequence; however, these results could also potentially be consistent with green
galaxies being a simple mix of the red and blue galaxy populations.

  \subsection{$\full$ ``Purple'' Galaxy Comparison Sample} \label{sec:purple} We
  first define a ``purple'' galaxy sample, which combines the red and blue
  galaxy populations to compare against the green galaxy population. Starting
  with the red and blue galaxy samples defined above, we weight the galaxies
  such that the combined sample has the same ratio of red to blue galaxies as
  galaxies in the green sample above and below the minimum in the color
  bimodality (the dashed line in Figure~\ref{fig:doublegauss} defining the
  center of the green valley). In this way, the ratio of galaxies above and
  below the minimum in the color bimodality is the same for the green galaxy
  sample and the comparison ``purple'' galaxy sample. We refer to this sample as
  the `full purple galaxy sample' (labeled \full in the figures below). The
  \full purple galaxy sample contains a total of 2,037 galaxies, weighted to an
  effective sample size of 487 galaxies, with a ratio of red to blue galaxies of
  0.82.

  \subsection{Removing Dust Obscured Galaxies}\label{sec:nuvr} While the optical
  color of galaxies reflects their star formation history and the age of their
  stellar populations, it can be influenced by the presence of dust. In
  particular, galaxies in the green and red populations may have older stellar
  populations than blue galaxies, or they may appear to be redder due to dust
  obscuration. As the goal of this paper is to study the morphologies of green
  galaxies as a potential transition population of galaxies that are moving from
  the blue cloud to the red sequence, we would like to identify and remove
  interlopers from the green and red populations that are dusty star forming
  galaxies.

  % -------------------- NUV-R Plot --------------------------------------------
  \begin{figure}
    \epsscale{1.1}
    \plotone{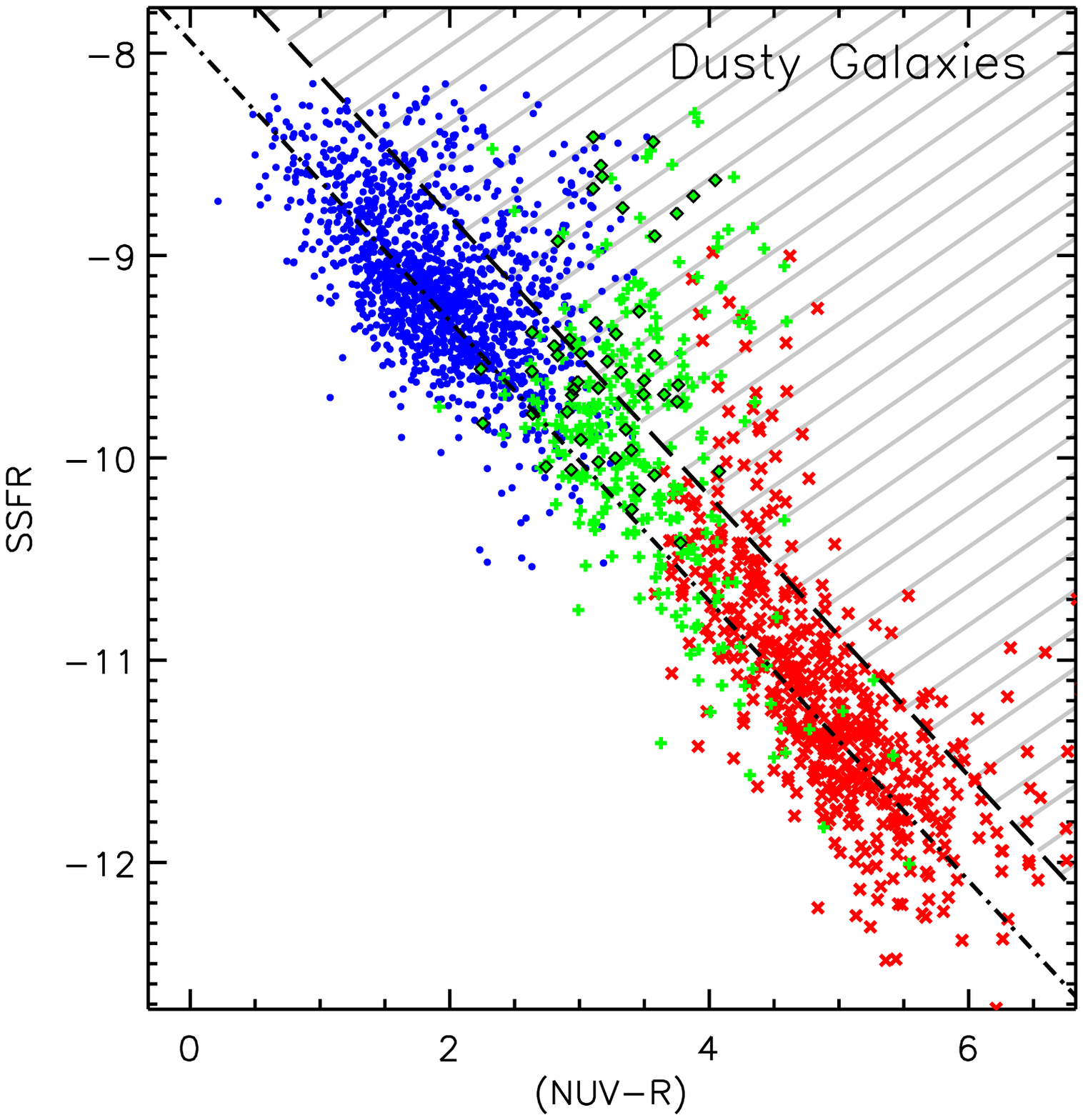}
    \caption{(NUV-R) color plotted versus specific star formation rate (SSFR)
    for the red, green, and blue galaxy samples used here. Galaxies with optical
    colors affected by dust will be scattered into the upper right corner of
    this plot, away from the main locus of galaxies (dot-dashed line). We
    identify and remove green and red dusty galaxies that lie more than
    $3\sigma$ (the dashed line) above this locus in our ``NUV'' samples.
    Green galaxies with $\bt=0$ are shown outlined by black diamonds.}
   \label{fig:nuvr}
  \end{figure}

  \citet{Salim09} show that a comparison between the SSFR versus rest-frame
  (NUV-R) color of $z\sim1$ galaxies (their Fig. 6) indicates that while the
  bulk of green galaxies (defined in \citet{Salim09} using (NUV-R) color) are
  transition objects with SSFRs intermediate between blue and red galaxies, a
  fraction of green galaxies are likely green due to dust in the host galaxy.
  These dusty green galaxies may therefore not be transition objects moving from
  the blue cloud to the red sequence. Following \citep{Salim09}, we identify and
  remove these dusty `interlopers' in our green and red galaxy samples. We use
  SSFR as derived in \citep{Salim09}, who use stellar population synthesis
  models to fit nine bands of photometry for AEGIS galaxies ($FUV$, $NUV$,
  $\ugriz$, $K_s$). Figure~\ref{fig:nuvr} shows SSFR versus (NUV-R) color for
  our optical color-selected galaxy samples. There is a relatively tight,
  well-defined locus in this space, as indicated by the dot-dashed line.
  Galaxies that lie $\ge3\sigma$ higher than this locus (in the upper right
  corner of Figure~\ref{fig:nuvr}, above the dashed line) are likely to have
  optical colors affected by dust obscuration. We remove these galaxies from the
  green and red samples used here to create a ``$\full\nuv$ green sample'' and a
  corresponding ``$\full\nuv$ purple sample'' that has the same ratio of red to
  blue galaxies (0.83) as the ``$\full\nuv$ green galaxy'' sample. All sample
  definitions denoted with subscript ``NUV'' have dusty obscured galaxies
  removed from the green and red populations. This procedure removes a total of
  123(36\%) green galaxies and 92(16\%) red galaxies. The $\full\nuv$ purple
  sample contains a total of 2003 galaxies, weighted to an effective sample size
  of 438 galaxies. The $\full\nuv$ green sample contains 219 galaxies.

  % ------------ Purple Distribution plot with scaled histograms ---------------
  \begin{figure*}
    \epsscale{0.9}
    \plotone{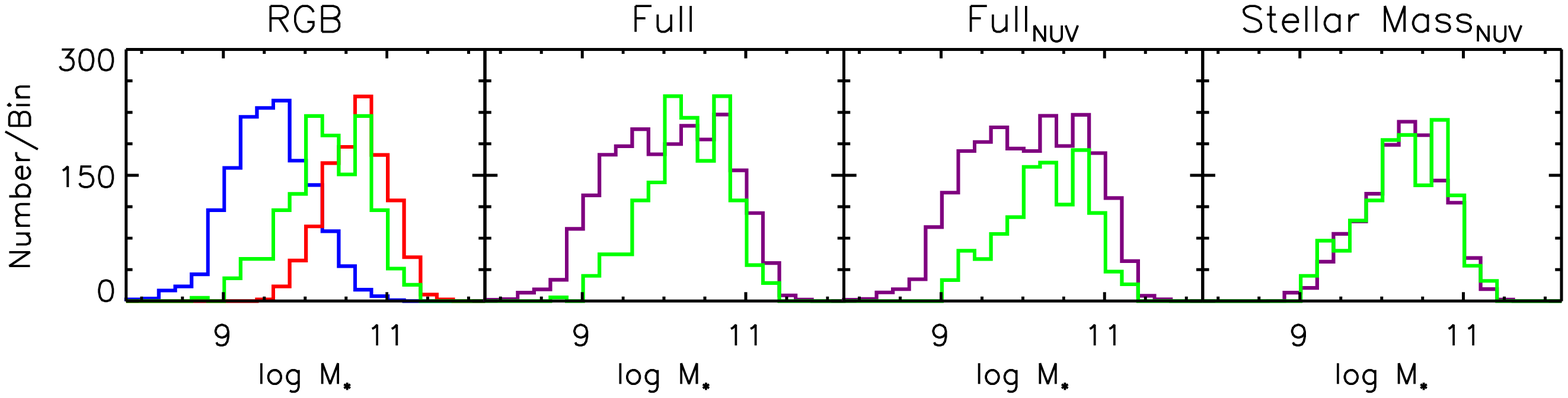}
    \caption{Stellar mass distributions for each galaxy sample. The left panel
    shows the red, green, and blue galaxy samples, while the other panels show
    various green and purple galaxy samples. For clarity we have weighted the
    histograms for red and blue galaxies by a factor of half and quarter,
    respectively.}
    \label{fig:stellarmass}
  \end{figure*}

  \subsection{Matched Stellar Mass Galaxy Samples} Figure~\ref{fig:stellarmass}
  shows the stellar mass distributions of the red, green, and blue galaxy
  samples (first panel), the $\full$ green and purple samples (second panel),
  and the $\full\nuv$ green and purple samples (third panel). As noted in
  Section~\ref{sec:properties}, optical color and stellar mass are highly
  correlated. As a result, the green galaxy samples have higher stellar mass, on
  average, than the corresponding purple comparison samples. To remove any
  dependence of morphology on stellar mass when comparing the green and purple
  samples, we create a purple comparison sample that has the same stellar mass
  distribution as the green $\full\nuv$ sample (shown in the right panel of
  Figure~\ref{fig:stellarmass}). To create this sample we assign weights to red
  and blue galaxies in the $\full\nuv$ purple sample in bins of ${\rm
  log}(M_{*}/M_{\sun}) = 0.1$ such that the stellar mass in each bin matches
  that of the stellar mass distribution of green galaxies either above (for red
  galaxies) or below (for blue galaxies) the minimum of the green valley. By
  weighting the red and blue galaxies to the green galaxies above and below the
  color minimum, we also constrain the ratio of red to blue galaxies to be the
  same as that within the green population (defined as above and below the
  minimum color that defines the center of the green valley) to within a few
  precent. The resulting $\stellarmass\nuv$ sample contains a total of 1970
  purple galaxies weighted to an effective sample size of 217 galaxies, with a
  red to blue ratio of 0.82.

  \subsection{``Purple'' vs Green Galaxy Comparison}\label{sec:purplesample}
  % Multicolor, CAS Plot -------------------------------------------------------
  \begin{figure*}
    \epsscale{1.}
    \plotone{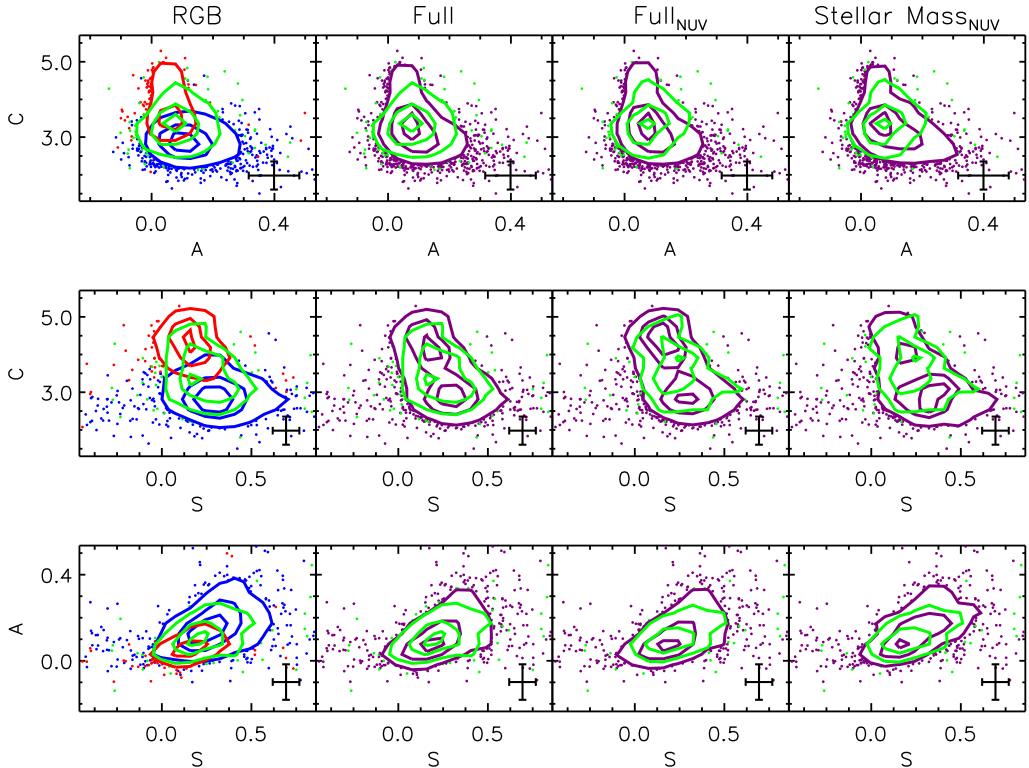}
    \caption{$\cas$ bivariate distributions shown for $\rgb$, $\full$, $\full\nuv$,
    and $\stellarmass\nuv$ samples. Contours are plotted to contain 30\%, 50\%
    and 80\% of the galaxies within each sample, with outliers shown as dots
    outside the 80\% contour. Estimates of 1 $\sigma$ error bars are shown in
    lower right corner for the median S/N pixel$^{-1}$ of the sample.}
    \label{fig:multicolorcas}
  \end{figure*}

  % Multicolor, GINI Plot ------------------------------------------------------
  \begin{figure*}
    \epsscale{1.}
    \plotone{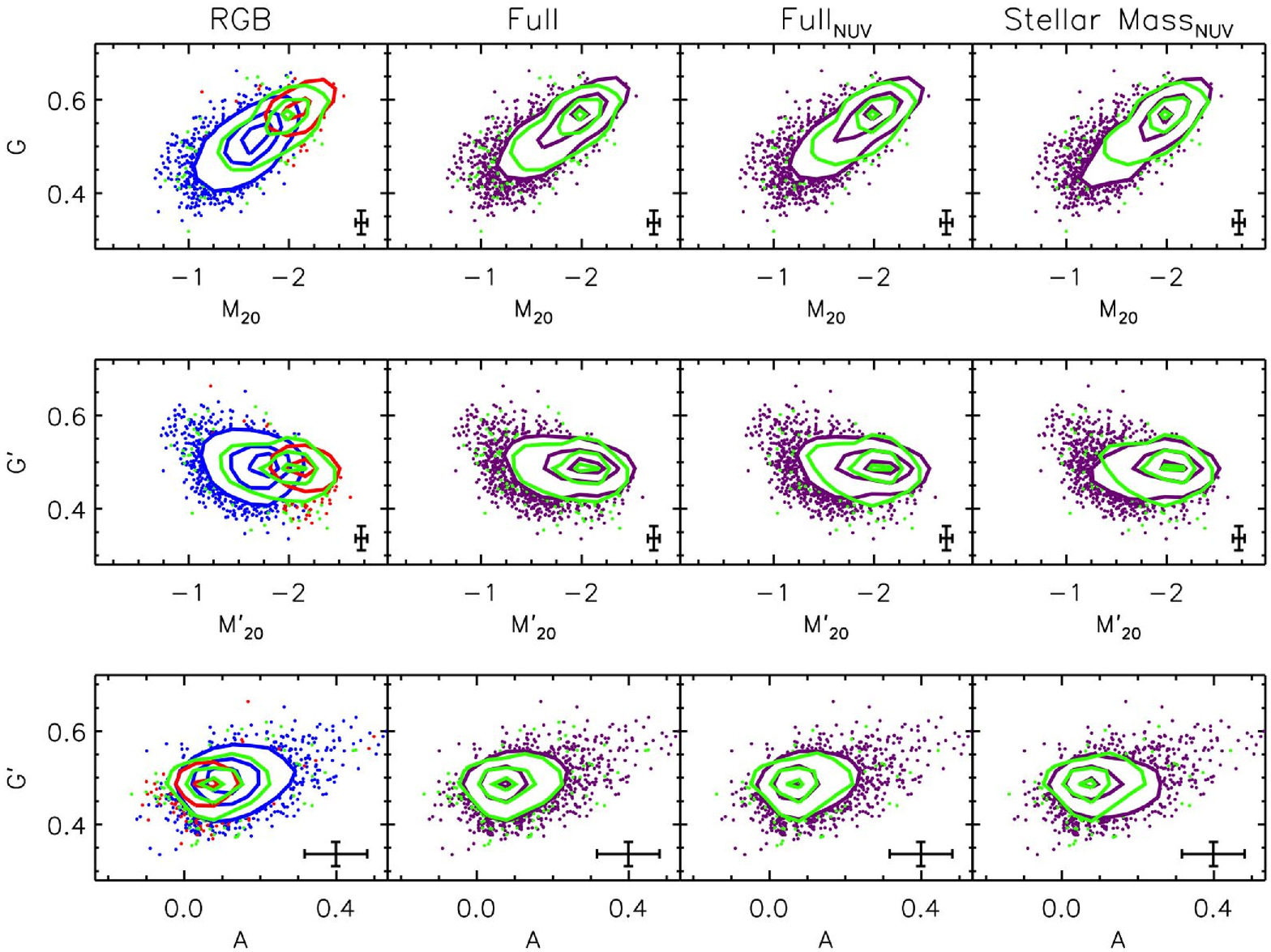}
    \caption{$G$-$\mtwe$, $\gtheta-\mtheta$, and $\gtheta$-A bivariate
    distributions shown for $\rgb$, $\full$, $\full\nuv$, and $\stellarmass\nuv$
    samples. Contours and error bars are similar to
    Figure~\ref{fig:multicolorcas}}
    \label{fig:multicolorgini}
  \end{figure*}

  % Multicolor, B.T Plot -------------------------------------------------------
  \begin{figure*}
    \epsscale{1.}
    \plotone{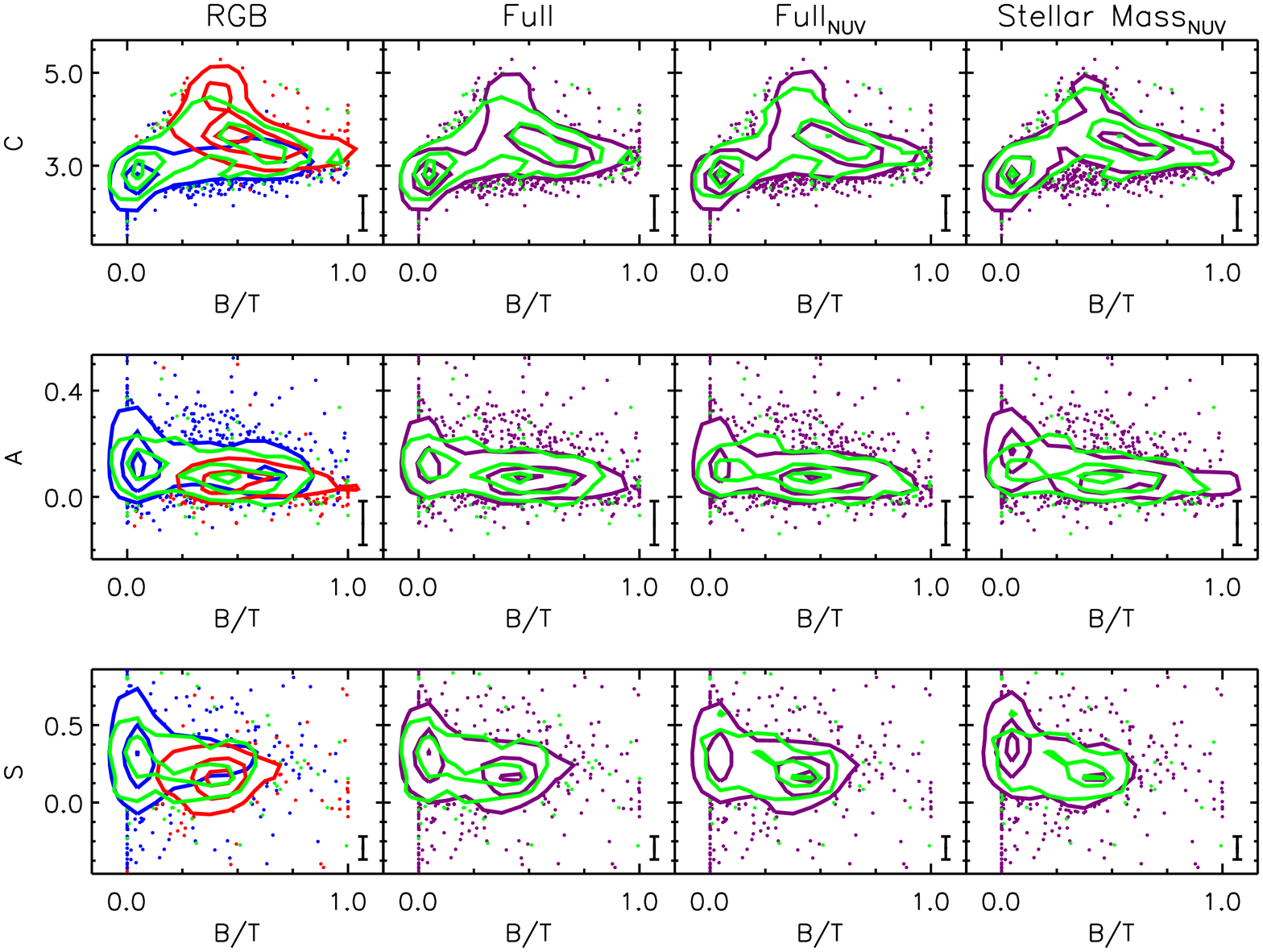}
    \caption{$\bt$-$\cas$ bivariate distributions shown for $\rgb$, $\full$,
    $\full\nuv$, and $\stellarmass\nuv$ samples. Contours and error bars are
    similar to Figure~\ref{fig:multicolorcas}}
    \label{fig:multicolorbt}
  \end{figure*}

  In Figures~\ref{fig:multicolorcas}-\ref{fig:multicolorbt} we plot
  distributions in $\cas$, $\bt$, and $\gtheta/\mtheta$ for the red, green and
  blue ($\rgb$) samples, alongside each of the green and purple comparison
  samples, $\full$, $\full\nuv$, and $\stellarmass\nuv$, to examine differences
  in the bivariate distributions of the green and purple galaxy samples. Contour
  levels and error bars are similar to Figure~\ref{fig:casparam}.

  In Figure~\ref{fig:multicolorcas} we compare the joint $\cas$ distributions of
  green and purple galaxies and find that the same trends that exist for the
  \full samples persist after we both remove dusty interlopers and match the
  samples in \stellarmass distribution. In concentration and asymmetry, we find
  a missing tail of high asymmetry ($A>0.2$) and low concentration ($C<3.0$)
  galaxies in the green samples, when compared to the purple samples.
  Additionally, the green galaxy population is also missing a tail of higher
  concentration galaxies ($C>4.5$) seen in the purple samples. These trends are
  reflected in the distributions of $C-S$ and $A-S$, where the green population
  is found to lack galaxies at the highest and lowest concentrations, as well as
  high asymmetry.

  In Figure~\ref{fig:multicolorgini} we show bivariate distributions of the
  green and purple comparison samples in the standard $G$/$\mtwe$ space as well
  as in the rotated $\gtheta-\mtheta$ space. We find that the green galaxy
  samples do not span the entire locus seen for the purple samples. In
  particular, the green galaxy samples do not contain objects with the lowest
  $G$ or $\mtwe$ values seen in the purple sample. This difference is particular
  clear in the $\stellarmass\nuv$ sample comparison. These differences are also
  seen in the rotated $\gtheta/\mtheta$ plane. We also compare the $\gtheta$ and
  asymmetry distributions, as both parameters are sensitive to mergers.
  Asymmetry is a tracer of major mergers, while $\gtheta$ tracers both major and
  minor mergers \citep{Lotz10}. We find that the green and purple samples have
  similar distributions in $\gtheta-A$. The $\stellarmass\nuv$ samples show a
  tail of galaxies to high $A$ values that is not seen in the green sample (as
  seen in Figure~\ref{fig:multicolorcas} above).

  We note that in Figures~\ref{fig:multicolorcas}-\ref{fig:multicolorbt} there
  are only minor differences between the $\full$, $\full\nuv$, and
  $\stellarmass\nuv$ samples, indicating that our results are not dominated by
  effects due to dust obscuration or the stellar mass-dependence of the green
  and purple galaxy samples.

  In Figure~\ref{fig:multicolorbt} we compare the distribution of $\bt$ with
  $C$, $A$, and $S$ for the green and purple samples. Here we find that the
  green galaxy population does not contain as many galaxies at high $\bt$ as the
  purple population, in addition to lacking galaxies at the highest asymmetry
  and smoothness values.

  % ------------------------ Multihistogram: KSTwo -----------------------------
  \begin{figure*}
    \epsscale{1.}
    \plotone{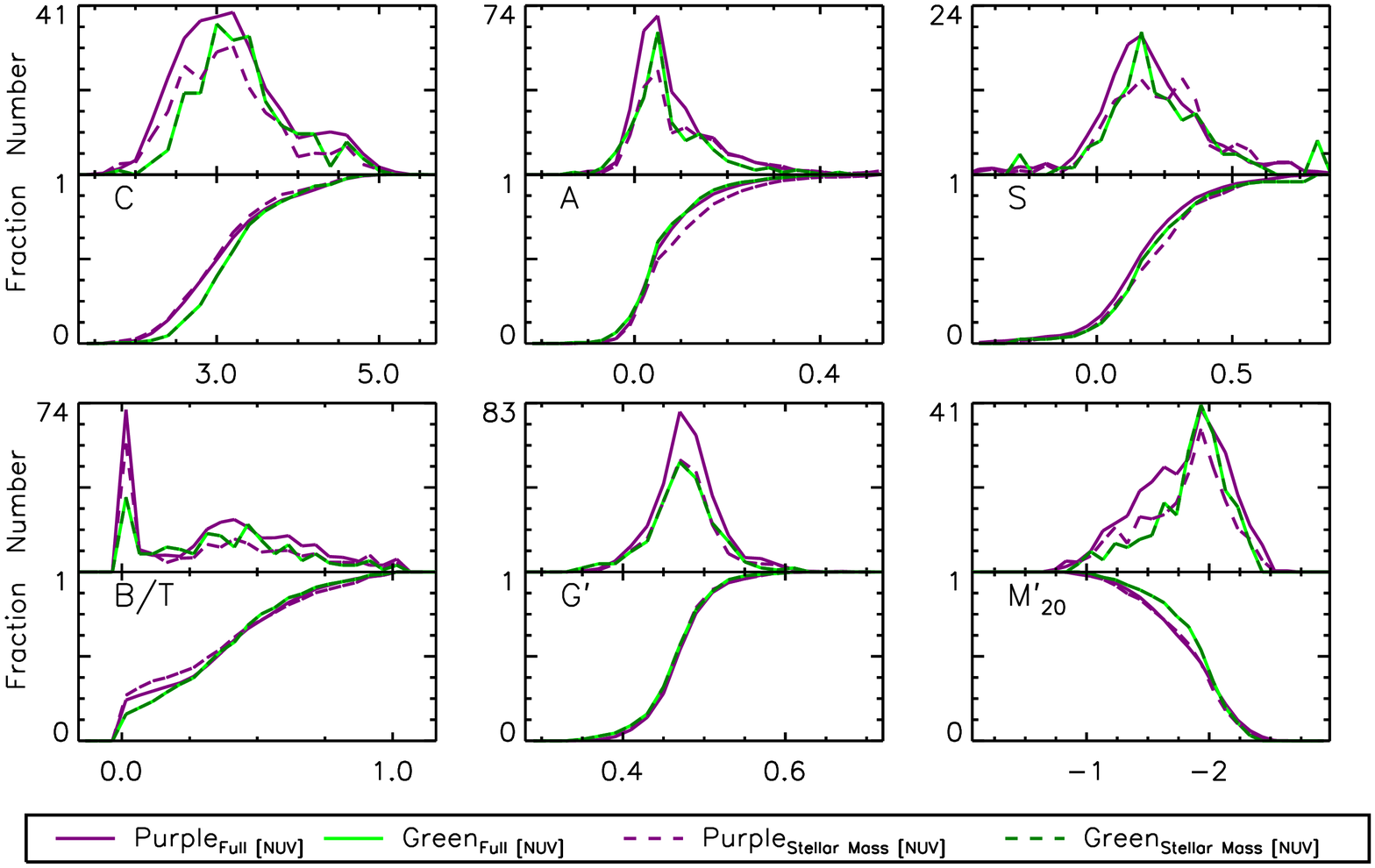}
    \caption{One-dimensional differential and cumulative distribution functions 
    for the Full$\nuv$ and Stellar Mass$\nuv$ green and purple galaxy samples
    in $C$, $A$, $S$, $\bt$, $\gtheta$, and $\mtheta$ space.  KS tests are 
    performed for each of these comparison samples; the results are given
    in Table~\ref{table:kstwo}.}
    \label{fig:multihistogram}
  \end{figure*}

  \subsection{Morphological Parameter Distribution Tests}\label{sec:kstests} For
  each of the measured morphological parameters, we apply the two-sided
  Kolmogorov-Smirnov (KS) statistic as a nonparametric null hypothesis test of
  the difference between the green and purple comparison samples. The null
  hypothesis is that both populations are drawn from the same parent sample. The
  KS statistic measures the maximal difference between the two normalized
  cumulative distribution functions and measures an associated significance
  level of rejecting the null hypothesis. In Figure~\ref{fig:multihistogram}, we
  plot the individual differential and cumulative distribution functions for
  each morphological parameter for the $\full\nuv$ and $\stellarmass\nuv$ green
  and purple samples. For the KS Statistic, if the probability is below either a
  1\% or 5\% significance level, we can reject the null-hypothesis that the two
  samples are drawn from a parent distribution at the 2 or 3 sigma levels,
  respectively.

  % KsTwo Table ----------------------------------------------------------------
  \begin{deluxetable}{rlrrr}  %Generated by textable.pro%
      \tablecaption{Kolmogorov-Smirnov two sample significance-levels \label{table:kstwo}\tablenotemark{a}}
      %SNIP%  10000 runs
      \tablehead{ & \colhead{Range} &                    \colhead{\full} &                \colhead{\full\nuv} &    \colhead{$\stellarmass\nuv$} \\[-1em]
 \hspace{0.625cm} &\hspace{0.625cm} &                    \hspace{1.25cm} &                    \hspace{1.25cm} &                  \hspace{1.25cm} }
      \startdata
		  log M$_{*}$ &     [7.8, 12.2] & \cellcolor[gray]{0.5}{ $<$ 0.01\%} & \cellcolor[gray]{0.5}{ $<$ 0.01\%} &                            72\% \\
		          $C$ &      [1.3, 5.7] &      \cellcolor[gray]{0.5}{0.32\%} &      \cellcolor[gray]{0.5}{0.16\%} &   \cellcolor[gray]{0.5}{0.73\%} \\
		          $A$ &     [-0.2, 0.5] &                               80\% &                               52\% &    \cellcolor[gray]{0.8}{3.5\%} \\
		          $S$ &     [-0.5, 0.9] &                               55\% &                               66\% &                            77\% \\
		          \bt &     [-0.2, 1.2] &       \cellcolor[gray]{0.8}{2.6\%} &                               12\% &    \cellcolor[gray]{0.8}{4.3\%} \\
		        \mtwe &    [-0.1, -3.0] &                              7.5\% &      \cellcolor[gray]{0.5}{0.39\%} &    \cellcolor[gray]{0.8}{2.0\%} \\
		          $G$ &      [0.3, 0.7] &                               43\% &                               72\% &                            25\% \\
		      \mtheta &    [-0.1, -3.0] &                              7.3\% &      \cellcolor[gray]{0.5}{0.41\%} &    \cellcolor[gray]{0.8}{1.9\%} \\
		      \gtheta &      [0.3, 0.7] &                               70\% &                               69\% &                            99\% \\
		    \rp [kpc] &    [-0.8, 15.8] &      \cellcolor[gray]{0.5}{0.34\%} &                               55\% &                            32\% \\
		  \enddata
      %SNIP%
      \tablenotetext{a}{KS test significance levels for each green and purple
      comparison sample. Parameters that can be rejected as being drawn from the
      same parent population at the 5\% and 1\% significance levels are
      highlighted with light grey and dark grey backgrounds, respectively.}
  \end{deluxetable}

  For the KS tests, we limit the effect of noisy measurements by applying the
  test on galaxies only within the parameter ranges shown in
  Figures~\ref{fig:casparam} and \ref{fig:giniparam} and listed in
  Table~\ref{table:kstwo}. We report the measured significance levels of the KS
  test results in Table~\ref{table:kstwo}. For each morphological parameter (as
  well as stellar mass and size), we compare the respective green and purple
  samples and highlight in Table~\ref{table:kstwo} the parameters that we can
  reject at the 5\% significance level in light grey and at the 1\% significance
  level in dark grey. Thus parameters highlighted in dark gray have therefore
  been rejected at the 3$\sigma$ level as being drawn from the same parent
  distribution. These include concentration for all of the $\full$, $\full\nuv$,
  and $\stellarmass\nuv$ comparison samples, as well as size ($\rp$) for the
  $\full$ sample and $\mtwe$ and $\mtheta$ for the $\full\nuv$ sample. At the
  2$\sigma$ level we can reject $\bt$ for the Full sample and A, $\bt$, $\mtwe$
  and $\mtheta$ for the $\stellarmass\nuv$ sample.

  We additionally performed KS tests to compare the morphological distributions
  of red and green, blue and green, and the union of red and blue to green
  samples, but we do not include them here as they are all rejected at the 1\%
  level. Smoothness, asymmetry, and the Gini coefficient in the union of red and
  blue compared with green sample tests were not rejected by the KS test.

  \subsection{Morphological Type Comparison using Gini/$\mtwe$}
  \label{sec:purpleginitype} We repeat the classification of galaxies into rough
  morphological types (early-type, late-type, or merger) using their location in
  $G$/$\mtwe$ space. Table~\ref{table:ginipurple} contains the fraction of green
  and purple galaxies in each sample associated with each morphological type. As
  before, error bars include both Poisson errors as well as Monte Carlo
  simulations of the scatter due to measurements errors of $G$/$\mtwe$. We do
  not find that the morphological type fractions are statistically different (at
  the 2$\sigma$ level) between any of the green and purple comparison samples.

%%%%%%%%%%%%%%%%%%%%%%%%%%%%%%%%%%%%%%%%%%%%%%%%%%%%%%%%%%%%%%%%%%%%%%%%%%%%%%%%
\section{Summary and Discussion} \label{sec:discussion}
%%%%%%%%%%%%%%%%%%%%%%%%%%%%%%%%%%%%%%%%%%%%%%%%%%%%%%%%%%%%%%%%%%%%%%%%%%%%%%%%
In this study we find that galaxies in the green valley at $0.4<z<1.2$ are an
intermediate population between galaxies in the blue cloud and red sequence in
terms of morphological type, concentration, asymmetry, smoothness, and merger
fraction. Our results do not change if we select green valley galaxies using
(NUV-R) color instead of (U-B) color. Using morphological types defined by
$G$/$\mtwe$, we find that the merger fraction ($14\%$), fraction of late-type
galaxies ($51\%$), and early-type galaxies ($35\%$) in the green valley to be
intermediate between the red and blue galaxy populations. We show that at a
given stellar mass, green galaxies have higher concentration values and lower
asymmetry and smoothness values than blue galaxies. They also have lower
concentration values and higher asymmetry and smoothness values than red
galaxies, at a given stellar mass. Additionally, 12\% of our green galaxy sample
is bulge-less, with $\bt=0$. Our results show that green galaxies are generally
massive ($M_{*}\sim 10^{10.5}$ $M_{\sun}$) disk galaxies with high
concentrations. Below we discuss the implications of these results.

  \subsection{Do green valley galaxies constitute a distinct population?} The
  idea that the bulk of the green valley galaxies are a transition population is
  clearly substantiated by our study of their morphological distribution
  compared to red sequence and blue cloud galaxies. While our optical color
  distribution is similarly fit by the two Gaussian distribution of
  \citet{Baldry04}, we find that galaxies in the green valley have significant
  morphological differences when compared to red and blue galaxies. This remains
  true if a UV-optical color selection is used instead.

  An important issue when interpreting the observed color bimodality of the CMD
  is that optical colors reflect not only the star formation history of a galaxy
  but also its dust content. The degeneracy between star formation history and
  dust can be partially corrected using the SED modeling \citep[particularly the
  UV color, e.g.][]{Salim07}, the Balmer decrement \citep[e.g.][]{Hopkins01,
  Brinchmann04}, or the IR to UV ratio \citep[e.g.][]{Buat05}. In contrast to
  claims that at higher redshifts there may not be a population of transition
  galaxies in the green valley \citep[e.g.][]{Brammer09}, we find that at
  $0.4<z<1.2$ in optically-selected samples there is a distinct population of
  galaxies that are green not due to dust obscuration but because they have an
  intermediate age stellar population \citep[see also][]{Salim09}. As we discuss
  in Section~\ref{sec:nuvr}, while some galaxies in our green valley sample are
  green from large SFRs and dust obscuration, our results do not change if they
  are removed from the sample.

  In addition to showing that green galaxies have different morphological
  distributions than either the red or blue galaxy populations alone, we further
  compare with control samples of joint red and blue galaxies with the same
  stellar mass. Our goal is to test whether the green galaxy population has a
  distinct distribution in any of our quantitative morphological measures than
  this joint set. This is difficult to test with the green galaxy sample being
  transition objects by nature and therefore having intermediate properties
  between the red and blue galaxy samples and is therefore a more stringent
  test. It is of course subject to errors and noise in the measured
  morphological parameters, and therefore a null result does not necessarily
  imply that there is not a difference; rather it may be hard to discern with
  these kinds of measurements. Additionally, the existence of other
  non-transition populations within the green galaxy population would only
  lessen the statistical differences detected between the green galaxy and
  comparison samples. However, we do find a 3$\sigma$ difference in terms of the
  distributions of concentration values and weaker 2$\sigma$ difference in the
  $\bt$, asymmetry, $\mtwe$, and $\gtheta$ distributions. Therefore it does
  appear that green galaxies must be a distinct population.

  % ------------------------ Gini Purple Type Table ----------------------------
  %$% Moved here so that latex would not allocate a page for this table.
  \begin{deluxetable*}{rrrrrrr}[b]  %Generated by textable.pro%
    \tablecolumns{7}
    \tablewidth{0pt}
    \tablecaption{Purple $G$/$\mtwe$ Morphological Types \label{table:ginipurple}\tablenotemark{a} }
    %SNIP%
    \tablehead{ &          \multicolumn{2}{c}{\full} &        \multicolumn{2}{c}{\full\nuv} &        \multicolumn{2}{c}{$\stellarmass\nuv$} \\
                & \colhead{Green} & \colhead{Purple} &  \colhead{Green} &  \colhead{Purple} &     \colhead{Green} &        \colhead{Purple}  }
    \startdata
	      Mergers &   14 $\pm$ 2 \% &    16 $\pm$ 2 \% &    15 $\pm$ 3 \% &     16 $\pm$ 2 \% &       15 $\pm$ 3 \% &           15 $\pm$ 3 \% \\
	   Early Type &   35 $\pm$ 4 \% &    34 $\pm$ 3 \% &    39 $\pm$ 5 \% &     36 $\pm$ 3 \% &       39 $\pm$ 5 \% &           34 $\pm$ 5 \% \\
	    Late Type &   51 $\pm$ 5 \% &    50 $\pm$ 4 \% &    46 $\pm$ 6 \% &     48 $\pm$ 4 \% &       46 $\pm$ 6 \% &           51 $\pm$ 6 \% \\
	  \enddata
    %SNIP%
    \tablenotetext{a}{Percents of each morphological type for each galaxy
    comparison sample, from $G$/$\mtwe$ classification for $0.4 < z \leq 1.2$.
    Uncertainties in each value are from Poisson errors and Monte Carlo
    Simulated Signal to Noise scatter using Median $G$/$\mtwe$ errors from
    \citet{Lotz06}.}
  \end{deluxetable*}

  \subsection{What mechanisms are quenching star formation in green valley
  galaxies?} Given that the green galaxy population is a transition population
  between the blue cloud and red sequence in which star formation was recently
  (within the last $\sim$ Gyr) quenched, it is the ideal population with which
  to study the quenching mechanism(s) at work. Merger-induced starbursts, which
  may quickly quench star formation, imprint high asymmetry and $\gtheta$ values
  on the galaxies. Importantly, we do \emph{not} find that most galaxies in the
  green valley are experiencing on-going mergers. In fact, the fraction of green
  galaxies identified as mergers using $G$/$\mtwe$ (and the fraction with high
  asymmetry, which is also often used to identify mergers) is \emph{lower} than
  the fraction within the blue galaxy population. If many of the green valley
  galaxies have recently undergone merger events, they must be at least
  $\sim$0.4 Gyr past the merger stage to not be identified using either
  $G$/$\mtwe$ or asymmetry \citep{Lotz10}. Furthermore, 51\% of the green
  galaxies in our sample are identified as late-type and either have not
  experienced a major merger recently or, if they did, it must have been gas
  rich \citep{Cox08}. However, using both quantitative measures to identify
  mergers, we find that the merger fraction is lower than that of blue galaxies
  and higher than that of red galaxies. While \citet{Kartaltepe10} find that
  most luminous infrared galaxies at $z\sim1$ are mergers, we do not find that
  most optically-selected galaxies at these redshifts, even those in the green
  valley, are identified as mergers.

  Additionally, we find that 12\% of green galaxies are bulge-less, with
  $\bt=0$. As a comparison, 30\% of blue galaxies and no red galaxies in our
  sample have $\bt=0$. The fact that 12\% of the green galaxies are bulge-less
  raises the question of how these green galaxies were created. Most($64\%$) of
  these galaxies are not green due to dust obscuration, as they have low SSFR
  indicative of intermediate age stellar populations. It is very unlikely that
  these galaxies have had major mergers in their recent past (which presumably
  would have built up a bulge component), and we have visually verified that
  they do not have central point sources indicative of an AGN, which could have
  biased the measured $\bt$ values. Somehow star formation must have been
  quenched in these objects without mechanisms that would also lead to the
  creation of a bulge.

  Higher concentration values seen in green galaxies could point to the presence
  of bars or other internal secular processes that slowly build a central bulge
  while allowing the disk structure to remain intact. In this scenario a
  gas-rich galaxy forms a stellar bar that funnels gas to the center of the
  galaxy, causing enhanced star formation and leading to the development of a
  bulge. This would increase the concentration measurement of these galaxies,
  though it does not explain the fraction of bulge-less green valley galaxies
  discussed above. \citet{Sheth05} find that there is clear evidence of more
  centrally-concentrated molecular gas distributions in barred spirals,
  supporting bar-driven transport of molecular gas to the central kiloparsec of
  galaxies. Barred spirals of late Hubble-types are less centrally concentrated
  than early Hubble-types, and there is enhanced star formation activity
  observed in early Hubble-type bars, indicating higher mass accretion rates
  \citep{Jogee05,Kormendy04}. Secular processes may also lead to some of the
  clumpiness that is seen in the profiles of green galaxies.

  The stellar disks of green galaxies do not appear to be truncated as they have
  similar, if not larger, sizes than blue galaxies, as indicated by their large
  Petrosian radii, which could be due to differences in their stellar mass
  distribution and the mass-size relationship \cite[e.g. ][]{Shen03} The tail
  of green galaxies with particularly large radii could reflect triggered star
  formation in the outer disks of these galaxies, possibly resulting from
  galaxy-galaxy interactions for a fraction of the sample. Green valley galaxies
  do not appear to be fading disks; the higher concentration values point to a
  different process. Fading disks would also produce higher $\bt$ values, which
  are not clearly seen here at a given stellar mass (though can not be ruled out
  by our data). This likely means that simple gas exhaustion is not the dominant
  mechanism.

  There may be a variety of quenching mechanisms responsible for the migration
  of galaxies from the blue cloud to the red sequence. While major mergers may
  play a role for some galaxies, we conclude that either a mild external
  process, such as galaxy harassment or tidal forces, or quite likely an
  internal process such as the creation of bars, is responsible for the
  quenching of star formation in many of green valley galaxies at $z\sim1$.

\vspace{2em}

%%% Support and Acknoledgements...----------------------------------------------
We thank our anonymous referee for useful comments that have greatly improved
this paper. We thank James Aird, TJ Cox, Aleks Diamond-Stanic, Jacqueline van
Gorkom, and Jim Gunn for helpful suggestions and discussions. We use data from
the DEEP2 survey, which was supported by NSF AST grants AST00-71048,
AST00-71198, AST05-07428, AST05-07483, AST08-07630, AST08-08133. This study
makes use of data from AEGIS Sruvey and in particular uses data from $\galex$,
$\hst$, Keck, and CFHT. The AEGIS Survey was supported in part by the NSF, NASA,
and the STFC. Based on observations obtained with MegaPrime/MegaCam, a joint
project of CFHT and CEA/DAPNIA, at the Canada-France-Hawaii Telescope (CFHT)
which is operated by the National Research Council (NRC) of Canada, the Institut
National des Science de l’Universof the Centre National de la Recherche
Scientifique (CNRS) of France, and the University of Hawaii. This work is based
in part on data products produced at TERAPIX and the Canadian Astronomy Data
Centre as part of the Canada-France-Hawaii Telescope Legacy Survey, a
collaborative project of NRC and CNRS.

\bibliography{references}

\end{document}